\documentclass[%
 aip,
 amsmath,amssymb,
preprint,%
author-year,%
]{revtex4-1}
\usepackage {CJK}
\usepackage{hyperref}
\hypersetup{
    colorlinks=true,
    linkcolor=blue,
    citecolor=blue,
    urlcolor=blue
}
\usepackage{graphicx}
\usepackage{dcolumn}
\usepackage{bm}
\usepackage[mathlines]{lineno}

\usepackage[utf8]{inputenc}
\usepackage[T1]{fontenc}
\usepackage{mathptmx}
\usepackage{etoolbox}

\makeatletter
\def\@email#1#2{%
 \endgroup
 \patchcmd{\titleblock@produce}
  {\frontmatter@RRAPformat}
  {\frontmatter@RRAPformat{\produce@RRAP{*#1\href{mailto:#2}{#2}}}\frontmatter@RRAPformat}
  {}{}
}%
\makeatother
\begin{document}

\preprint{AIP/123-QED}

\title[]{Symmetry breaking of rotating convection due to Non-Oberbeck--Boussinesq effects}
\author{Shuang Wang}
 \email{shuangw@mit.edu}
 \altaffiliation{Earth, Atmospheric and Planetary Science Department, Massachusetts Institute of Technology, Cambridge, MA 02139, USA}
\author{Wanying Kang}%
 \affiliation{Earth, Atmospheric and Planetary Science Department, Massachusetts Institute of Technology, Cambridge, MA 02139, USA}%

\date{\today}

\begin{abstract}
The non-Oberbeck--Boussinesq (NOB) effects arising from variations in thermal expansivity are theoretically and numerically studied in the context of rotating Rayleigh--B\'{e}nard convection in forms of two-dimensional (2D) rolls. The thermal expansivity increases with pressure (depth), and its variation is measured by a dimensionless factor $\epsilon$. Utilizing an asymptotic expansion with weak nonlinearity, we derive an amplitude equation, revealing that NOB effects amplify the magnitude of convection. An $\epsilon^2$-order NOB correction leads to a symmetry breaking about the horizontal mid-plane, manifested in the strengthening of convection near the bottom and its weakening near the top, forming bottom-heavy profiles. At $\epsilon^3$-order, the conjunction of NOB effects and nonlinear advection leads to a horizontal symmetry breaking. The values of Taylor number and Prandlt number determine whether upward or downward plumes are stronger. Numerical calculations validate the theoretical analyses in weakly nonlinear regime. This work advances our understanding of hydrothermal plumes in some winter lakes on Earth, and in the subglacial oceans on icy moons as well as tracer transport from the seafloor to the ice shell. 
\end{abstract}

\maketitle

\section{Introduction}
Rayleigh--B\'{e}nard convection (RBC) commonly exists in nature, manifested in various scenarios such as afternoon convection in the atmosphere, hydrothermal plumes on the seafloor, and flows in the convective zone of the sun. The RBC problem is generally investigated under the Oberbeck--Boussinesq (OB) approximation, where fluid properties are assumed to be constant, except for a linear relation between density and temperature in the buoyancy term \citep{Oberbeck_1879,Boussinesq_1903}. However, this approximation becomes invalid when flow compressibility or variations of fluid properties become non-negligible, as demonstrated by \cite{Gray_and_Giorgini_1976}. Departures from the OB approximation are called non-Oberbeck--Boussinesq (NOB) effects. Changes of fluid properties commonly exist in convective systems in the nature, especially when the temperature and pressure differences are large \citep[e.g.,][]{Zhang_et_al_1997,Ahlers_et_al_2006,Ahlers_et_al_2008}. However, sometimes, even small temperature differences can lead to notifiable NOB effects because fluid properties are very sensitive to the changes of temperature and pressure. As an example, fresh water's thermal expansivity rapidly varies with temperature and can change sign near freezing point \citep{Wang_et_al_2019}.     

NOB effects arising from variations in fluid properties under large temperature differences in non-rotating convection have been extensively investigated. Many previous works have found NOB-induced symmetry breaking about the horizontal mid-plane in temperature drops and the thicknesses of boundary layers, in water \citep[e.g,][]{Ahlers_et_al_2006,Sugiyama_et_al_2009,Horn_Shishkina_2014}, glycerol \citep[e.g.,][]{Zhang_et_al_1997,Sugiyama_et_al_2007}, gases such as ethane \citep[e.g.,][]{Ahlers_et_al_2007,Ahlers_et_al_2008} and air \citep[e.g.,][]{Liu_et_al_2018, Wan_et_al_2020}. This symmetry breaking is related to a shift of the bulk temperature of the convective cell away from the corresponding temperature in the OB convection, due to the difference in heat/momentum transport efficiency across the top and bottom boundary layers as fluid diffusivity and viscosity vary. Specifically, a decrease in viscosity with temperature has been shown to be able to enhance mixing near the cold surface relative to the hot surface, thereby raising the bulk temperature and breaking the vertical symmetry in water convection \citep{Ahlers_et_al_2006, Horn_Shishkina_2014, Sameen_2009}. This asymmetry can be well explained by generalizing Prandtl--Blasius boundary-layer theory to account for variations in viscosity \citep{Ahlers_et_al_2006}. Applying the same theory to ethane convection fails because the variation of thermal expansivity dominates \citep{Ahlers_et_al_2007,Ahlers_et_al_2008, Sameen_2009}. As temperature drop starts to become vertically asymmetric, the fluid motion also changes. \cite{Busse_1967} theoretically investigated NOB effects on RBC pattern near onset, revealing an asymmetric hexagonal pattern with a stronger up/down-flow center and weaker downward/upward side-flows. This pattern was further confirmed experimentally and theoretically by \cite{Ahlers_et_al_2010} using $\rm SF_6$ near its gas–liquid critical point. Under two-dimensional configuration, \cite{Liu_et_al_2018} analytically and numerically demonstrated the asymmetry of motion in convective cells. 

Since convection in meteorology, geophysics, and astrophysics is commonly influenced by the rotation of planets and stars, it is essential to understand the behavior of a rotating RBC system \citep[e.g.,][]{Chandrasekhar_1953, Scheel_2007, Kunnen_2021}. While previous literature extensively investigates the effects of rotation on the convection under the OB approximation, research remains relatively limited under the NOB scenario. \cite{Horn_Shishkina_2014} explored NOB effects caused by variations in viscosity, diffusivity and thermal expansivity in rotating RBC and observed that strong rotation suppresses NOB effects, evident in a sharp weakening in the shift of the bulk temperature of the convective cell. Nevertheless, there is still a pressing need for systematic investigations into convection in the presence of both rotation and NOB effects, as emphasized in the review by \cite{Ecke_and_Shishkina_2023}.   

\begin{figure*}
    \centering
    \includegraphics[width=0.8\textwidth]{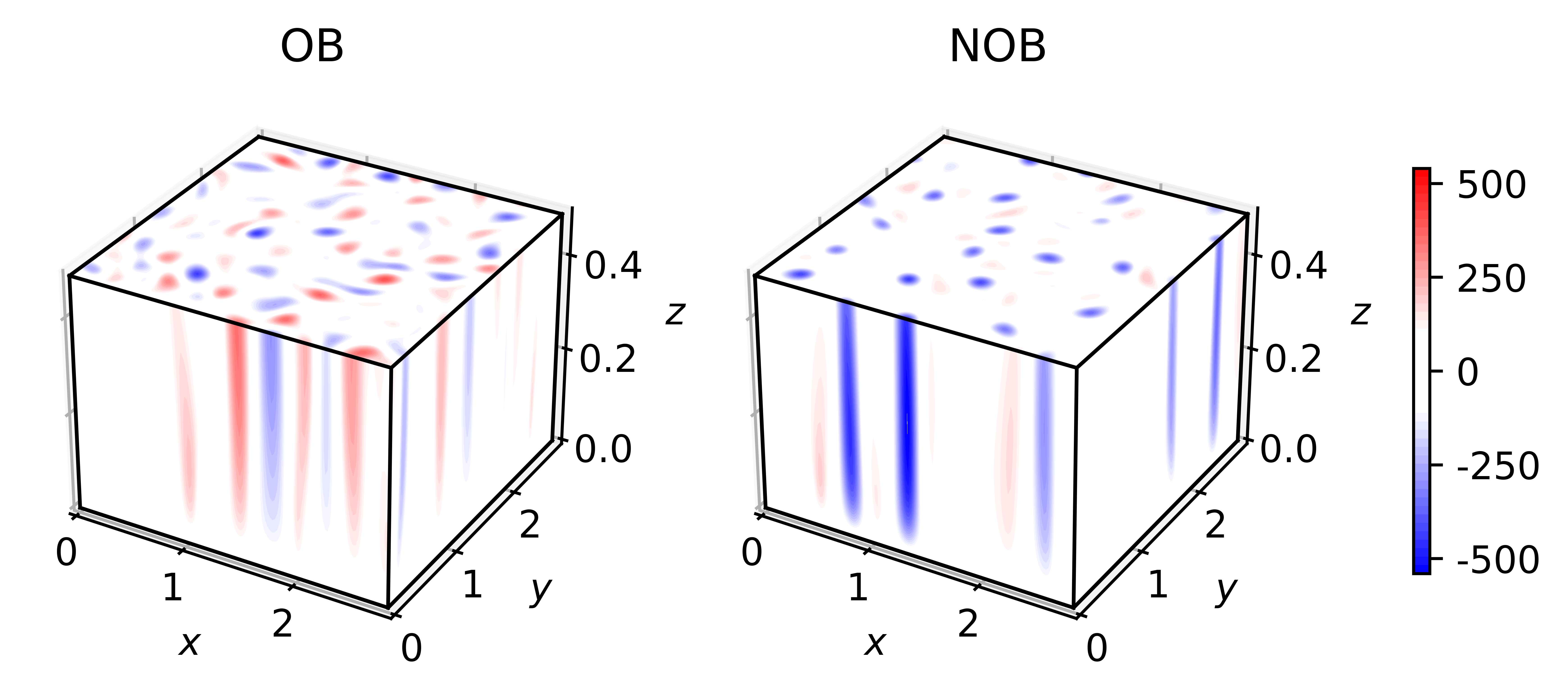}
    \caption{Dimensionless vertical velocities in the lower part of the domain ($0<z<0.5$) in three-dimensional simulations in OB and NOB scenario. Rayleigh number is $R_a=10^6$, Talyor number is $T_a=10^6$, and Prandlt number is $P_r=7$. The numerical solver is Dedalus. The dimensionless size of the domain is $2\sqrt{2}\times2\sqrt{2}\times1$, and the resolution is $256\times256\times128$.}
    \label{fig:3d_case}
\end{figure*}

In this paper, we use asymptotic expansion and numerical simulations to investigate NOB effects induced by the change of thermal expansivity $\hat{\alpha}$ in a 2D rotating RBC system, motivated by convection in winter lakes on Earth \citep[e.g.,][]{Wuest_et_al_2005,Bouffard_et_al_2016,Bouffard_and_Wuest_2019} and in subglacial oceans on icy moons \citep[e.g.,][]{Melosh_et_al_2004,Zeng_and_Jansen_2021,Kang_et_al_2022}. In presence of bottom heating, temperature and pressure should increase with depth, causing $\hat{\alpha}$ to increase with depth, following fresh water's equation of state. While other fluid properties, such as viscosity and diffusivity also vary with depth, their changes play a negligible role compared to the influence of $\hat{\alpha}$ variations. This is particularly true near water's freezing point, where $\hat{\alpha}$ is close to zero and can change sign as temperature and pressure vary \citep{Mallamace_et_al_2013}.  Recently, \cite{Kang_et_al_2022} used an oceanic general circulation model to simulate convective plumes on Enceladus (a moon of Saturn), taking into account this type of NOB effects. They found convection to be dominated by downward cold plumes, even though the ocean is mainly heated from the bottom (replicated in Fig.\ref{fig:3d_case}). With all strong plumes oriented downward, tracers injected by hydrothermal vents on the seafloor may be much harder to be transported upward to encounter the ice shell. 
The structure of this paper is as follows. Problem setup is provided in Sec.~\ref{sec:method}. The major theoretical analyses based on linear and weakly nonlinear analysis are presented in Sec.~\ref{sec:instability}. Numerical simulations and results are displayed in Sec.~\ref{sec:numerical_results}. Section~\ref{sec:summary} is a brief summary and discussion.

\section{Problem setup}\label{sec:method}
We study rotating RBC system, where the fluid is confined between a cold top plate and a warm bottom plate, separated by $\hat{H}$ (hatted quantities are dimensional). The two plates are thermostatic, with temperatures fixed at $\hat{\theta}^t$ and $\hat{\theta}^b$, respectively ($\hat{\theta}^b>\hat{\theta}^t$). The horizontal domain is assumed to be infinitely large, and the system rotates around the vertical axis. The governing equations are as follows
\begin{eqnarray}
    \frac{\partial\hat{\bf{u}}}{\partial\hat{t}}&=&-\hat{{\bf u}}\cdot\hat{\nabla}\hat{{\bf u}}-\hat{f}{\bf k}\times\hat{{\bf u}}-\hat{\nabla}\hat{\phi}+\hat{\nu}\hat{\nabla}^2\hat{{\bf u}}+\hat{g}\hat{\alpha}\hat{\theta}{\bf k}, \\
    \frac{\partial\hat{\theta}}{\partial\hat{t}}&=&-\hat{\bf{u}}\cdot\hat{\nabla}\hat{\theta}+\hat{\kappa}\hat{\nabla}^2\hat{\theta}, \\
    \frac{\partial\hat{w}}{\partial\hat{z}}&=&-\left(\frac{\partial\hat{u}}{\partial\hat{x}}+\frac{\partial\hat{v}}{\partial\hat{y}}\right),
\end{eqnarray}
where $({\bf i},\,{\bf j},\,{\bf k})$ represent unit vectors along $x-$, $y-$, and $z-$axes, $\hat{{\bf u}}=(\hat{u},\,\hat{v},\,\hat{w})$ denotes the velocity components along axes, $\hat{\phi}$ ($=\hat{p}/\hat{\rho}$, $\hat{p}$ is hydrodynamic pressure and $\hat{\rho}$ is fluid density) is geopotential, $\hat{\theta}$ is temperature, $\hat{f}$ is the Coriolis parameter, $\hat{g}$ is the gravity, $\hat{\nu}$ is constant viscosity, $\hat{\kappa}$ is constant diffusivity, and $\hat{\alpha}$ is thermal expansivity. For convenience, we non-dimensionalize these equations into  
\begin{eqnarray}
    \frac{1}{P_r}\frac{\partial{\bf u}}{\partial t}&=&-\frac{1}{P_r}{\bf u}\cdot\nabla{\bf u}-\sqrt{T_a}{\bf k}\times{\bf u}-\nabla\phi+\nabla^2{\bf u}+R_a[1+\epsilon\cos{(\pi z)}]\theta{\bf k}, \label{eq:general_du/dt} \\
    \frac{\partial\theta}{\partial t}&=&-{\bf u}\cdot\nabla\theta+\nabla^2\theta, \label{eq:general_dtheta/dt} \\
    \frac{\partial w}{\partial z}&=&-\left(\frac{\partial u}{\partial x}+\frac{\partial v}{\partial y}\right), \label{eq:general_mass_conserve}
\end{eqnarray}
where the temperature difference between the top and bottom plates $\Delta\hat{\theta}$, domain's height $\hat{H}$, reference velocity $\hat{U}=\hat{\kappa}/\hat{H}$ ($\hat{\kappa}$ is constant diffusivity of the fluid), reference time $\hat{H}/\hat{U}$, mean density $\hat{\rho}_0$, and reference hydrodynamic pressure $\hat{p}_0=\hat{\rho}_0\hat{U}^2$ are utilized. The three dimensionless parameters in the above equations include the Rayleigh number $R_a$, the Taylor number $T_a$, and the Prandlt number $P_r$, defined as follows:
\refstepcounter{equation}
$$
  R_a=\frac{\hat{g}\hat{\alpha}_0\hat{H}^3\Delta\hat{\theta}}{\hat{\kappa}\hat{\nu}},\quad T_a=\frac{\hat{f}^2\hat{H}^4}{\hat{\nu}^2},\quad P_r=\frac{\hat{\nu}}{\hat{\kappa}},
  \eqno{( \theequation{\mathit{a},\mathit{b},\mathit{c}})}
$$
where $\hat{\alpha}_0$ is the mean thermal expansivity of the fluid. Besides, an extra dimensionless parameter, $\epsilon$, is defined to measure the strength of NOB effects arising from the variation of thermal expansivity, i.e., $\epsilon=\Delta\hat{\alpha}/(2\hat{\alpha}_0)$, where $\Delta\hat{\alpha}$ is the difference in thermal expansivity between the top and bottom plates.

It is important to acknowledge that the profile of the variation in $\hat{\alpha}$ used here is $\hat{\alpha}_0\cos{(\pi z)}$, and then manifest in an additional term $\epsilon R_a\cos{(\pi z)}$ in Eq.~(\ref{eq:general_du/dt}). It increases with depth and the cosine dependence allows us to do asymptotic expansion more easily. However, realistic $\hat{\alpha}$ profiles can be more complicated. For instance, in a freshwater subglacial ocean on an icy moon, $\hat{\alpha}$ might rapidly increase towards greater depth in a shallow upper layer and almost remain a constant in deeper ocean, connected by a sharp transition layer \citep[e.g.,][]{Zeng_and_Jansen_2021}. Despite the difference in the detailed $\hat{\alpha}$ profile, the qualitative results achieved here should still provide useful insights, as long as $\hat{\alpha}$ increases with depth.

The governing equations are equivalent to the vertical vorticity equation
\begin{equation}\label{eq:general_dvort/dt}
    D_{\nu}q=-\sqrt{T_a}\sigma+\frac{1}{P_r}[\nabla\times(-{\bf u}\cdot\nabla{\bf u})]\cdot{\bf k},
\end{equation}
the horizontal divergence equation
\begin{equation}\label{eq:general_dconv/dt}
    D_{\nu}\sigma=\sqrt{T_a}q-\nabla_h^2\phi+\frac{1}{P_r}\nabla_h\cdot(-{\bf u}\cdot\nabla{\bf u}), 
\end{equation}
and the vertical velocity equation
\begin{eqnarray}\label{eq:general_rRBC_w}
    D_{\kappa} 
    \left(D_{\nu}^2\nabla^2+T_a\frac{\partial^2}{\partial z^2}\right)w&=&D_{\kappa}D_{\nu}\{R_a[1+\epsilon\cos{(\pi z})]\nabla_h^2\theta\}-\frac{\sqrt{T_a}}{P_r}D_{\kappa}\frac{\partial}{\partial z}[\nabla\times(-{\bf u}\cdot\nabla{\bf u})]\cdot{\bf k} \nonumber \\
    &&-\frac{1}{P_r}D_{\kappa}D_{\nu}\frac{\partial}{\partial z}\nabla_h\cdot(-{\bf u}\cdot\nabla{\bf u})+\frac{1}{P_r}D_{\kappa}D_{\nu}\nabla^2_h\left(-{\bf u}\cdot\nabla w\right),
\end{eqnarray}
where 
\begin{subequations}
    \begin{eqnarray}
        D_{\nu}&=&\frac{1}{P_r}\frac{\partial}{\partial t}-\nabla^2,\quad\quad D_{\kappa}=\frac{\partial}{\partial t}-\nabla^2, \\
        q&=&\frac{\partial v}{\partial x}-\frac{\partial u}{\partial y},\qquad\quad\,\,\,\sigma=\frac{\partial u}{\partial x}+\frac{\partial v}{\partial y}, \\
        \nabla_h&=&{\bf i}\frac{\partial}{\partial x}+{\bf j}\frac{\partial}{\partial y},\qquad\,\,\,\, \nabla_h^2=\nabla_h\cdot\nabla_h.
    \end{eqnarray}
\end{subequations}

Although the previous equations are generally applicable to both 2D and 3D systems, here onwards, we focus on the 2D configuration, which allows us to simplify the asymptotic analysis and to access a wider parameter space in numerical simulations.  In this 2D configuration, all fields are invariant in $y$ and only vary with $x$ and $z$.


The equations are constrained by the following thermostatic and stress-free boundary conditions:
\begin{subequations}\label{eq:boundary_condition}
    \begin{eqnarray}
        \theta&=&\theta^b,\quad\quad\,\, \mbox{for}\quad z=0, \\
        \theta&=&\theta^t,\quad\quad\,\,\,\mbox{for}\quad z=1, \\
        w&=&\nabla^2_zw=0,\quad\,\,\,\,\,\mbox{for}\quad z=0,\,1.
    \end{eqnarray} 
\end{subequations}

\section{Linear and weakly nonlinear analysis}\label{sec:instability}
We follow the procedure outlined in \cite{Scheel_2007} and \cite{Liu_et_al_2018} to do linear and weakly nonlinear analysis, assuming all quantities are a superposition of the following asymptotic expansions in powers of $\epsilon$:
\begin{equation}\label{eq:series}
    \mathcal{A}= \mathcal{A}_0+\epsilon\mathcal{A}_1+\epsilon^2\mathcal{A}_2+\epsilon^3\mathcal{A}_3+..., \nonumber
\end{equation}
where $\mathcal{A}$ denotes $({\bf u},\,\theta,\,q,\sigma,\,\phi)$. Since Eqs. (\ref{eq:general_du/dt})--(\ref{eq:general_rRBC_w}) can be written in a power series of $\epsilon$. The spatial structure and amplitude of the state variables are obtained by solving the reduced set of governing equations at only a specific order of $\epsilon$.

\subsection{Linear instability analysis}
Before the onset of convection, a quiescent conductive state is a stable solution. The governing equations are reduced to
\begin{eqnarray}
    \frac{\mathrm{d}^2\theta_0}{\mathrm{d}z^2}&=&0, \\
    \frac{\mathrm{d}\phi_0}{\mathrm{d}z}&=&R_a\theta_0.
\end{eqnarray}
The corresponding solutions are
\begin{subequations}
    \begin{eqnarray}
        u_0&=&v_0=w_0=q_0=\sigma_0=0, \\
        \theta_0&=&(\theta^t-\theta^b)z+\theta^b, \\
        \phi_0&=&R_a\int_0^z\theta_0\mathrm{d}z'+\epsilon R_a \int_0^z\theta_0\cos{(\pi z')}\mathrm{d}z'.
    \end{eqnarray}
\end{subequations}
Note that the second term on the right-hand side of $\phi_0$ is retained despite its dependence on the power of $\epsilon$, as it only appears in $\phi_0$ and does not generate any motion.

The motions of convection occur at $\epsilon$-order and higher orders. At $\epsilon$-order, Eq. (\ref{eq:general_rRBC_w}) has the form of
\begin{equation}\label{eq:1st_rRBC_w}
    \left[D_{\kappa} \left(D_{\nu}^2\nabla^2+T_a\frac{\partial^2}{\partial z^2}\right)-R_a D_{\nu}\nabla_h^2\right]w_1=0.
\end{equation}
The simplest solution of $w_1$ satisfying the boundary conditions (\ref{eq:boundary_condition}) is a two-dimensional roll described as
\begin{equation}
    w_1\sim e^{s t}\sin{(kx+ly)}\sin{mz},
\end{equation}
where $s$ is the growth rate, $k$, $l$, and $m$ are the wavenumbers in $x$-, $y$-, and $z$-directions. Substituting this solution into Eq. (\ref{eq:1st_rRBC_w}) leads to the dispersion relation
\begin{equation}\label{eq:1st_dispersion_relation}
    s^3+Bs^2+Cs+D=0,
\end{equation}
where 
\begin{subequations}
    \begin{eqnarray}
        B&=&(2P_r+1)K_{11}^2, \\
    C&=&P_r(P_r+2)K_{11}^4+P_r^2T_a\frac{m^2}{K_{11}^2}-P_rR_a\frac{k_h^2}{K_{11}^2}, \\
    D&=&P_r^2(K_{11}^6+T_am^2-R_ak_h^2).
    \end{eqnarray}
\end{subequations}
Here, $k_h^2=k^2+l^2$ and $K_{11}^2=k_h^2+m^2$. Due to the presence of rotation, two marginal instabilities can occur. One is a steady convection corresponding to $D=0$, and the other is an oscillatory state corresponding to $BC=D$ \citep{Chandrasekhar_1953,Boubnov_and_Golitsyn_1995,Wood_and_Bushby_2016,Yano_2023_chap5}. In the present paper, we solely focus on the onset of the first instability at $D=0$, which yields a critical Rayleigh number 
\begin{equation}\label{eq:critical_Rayleigh_number}
    R_{ac}=\frac{K_{11}^6+T_a m^2}{k_h^2}=\frac{\hat{g}\hat{\alpha}_0\hat{H}^3\Delta\hat{\theta}_c}{\hat{\kappa}\hat{\nu}},
\end{equation}
where $\Delta\hat{\theta}_c$ is the critical temperature difference between the top and bottom plates. The $R_{ac}$ above does not involve any NOB correction, and is identical to the OB case. It is also evident that rotation postpones the onset of convection, as manifested by an increase in $R_{ac}$ with $T_a$. In the rapidly rotating regime, $R_{ac}$ from linear analysis well matches the two-third power law of $8.7T_a^{2/3}$ proposed by \cite{Chandrasekhar_1953}, where the factor 8.7 is a numerical approximation at very large $T_a$. 

The vertical wavenumber of the gravest mode is $m=\pi$. Also, the horizontal wavenumber of the gravest mode should satisfy $\partial R_{ac}/\partial k_h^2=0$, leading to
\begin{equation}\label{eq:critical_1st_wavenumber}
    2\left(\frac{k_h^2}{m^2}\right)^3+3\left(\frac{k_h^2}{m^2}\right)^2-1=\frac{T_a}{m^4}=\frac{T_a}{\pi^4}.
\end{equation}
The cubic function on the left-hand side monotonically increases as $k_h^2/m^2$ increases. The truncation condition is $k_h^2/m^2\geq 1/2$ due to the constraint $T_a\geq 0$. 

Following \cite{Liu_et_al_2018}, we study the weakly nonlinear regime where the supercriticality $\delta R_a/R_a=(R_a-R_{ac})/R_{ac}$ is $\epsilon^2$-order. By so doing, convection is mostly in a steady state with its amplitude slowly varying with time. Details about the factors that control the convection amplitude is discussed in Sec.~\ref{subsec:amplitude}.

\subsection{Asymptotic solution}\label{sec:pattern}
At $\epsilon$-order, we obtain the structure of the most unstable mode
\begin{subequations}
    \begin{eqnarray}
        u_1&=&U_1\cos{(kx+ly)}\cos{mz}, \\
        v_1&=&V_1\cos{(kx+ly)}\cos{mz}, \\
        w_1&=&W_1\sin{(kx+ly)}\sin{mz}, \\
        \theta_1&=&\Theta_1\sin{(kx+ly)}\sin{mz}, \\
        q_{1}&=&Q_1\sin{(kx+ly)}\cos{mz}, \\
        \sigma_{1}&=&\Sigma_1\sin{(kx+ly)}\cos{mz}.
    \end{eqnarray}
\end{subequations}
Coefficients in the above equations are given in Appendix \ref{sec:appA}. As depicted in Fig.~\ref{fig:mechanism_diagram}(\emph{a}), the linear convection solution features a pair of upward and downward plume. Since NOB correction has not entered the equation at $\epsilon$-order, the solution is identical to the OB case. 


Nonlinear advection and NOB effects together lead to a correction at the $\epsilon^2$-order,
\begin{subequations}
    \begin{eqnarray}
        u_2&=& U_{21}\cos{(kx+ly)}\cos{2mz}+U_{22}\sin{2(kx+ly)}, \\
        v_2&=& V_{21}\cos{(kx+ly)}\cos{2mz}+V_{22}\sin{2(kx+ly)}, \\
        w_2&=&W_2\sin{(kx+ly)}\sin{2mz}, \\
        \theta_2&=& \Theta_{21}\sin{(kx+ly)}\sin{2mz}+\Theta_{22}\sin{2mz}, \\
        q_{2}&=& Q_{21}\sin{(kx+ly)}\cos{2mz}+Q_{22}\cos{2(kx+ly)}, \\
        \sigma_{2}&=& \Sigma_2\sin{(kx+ly)}\cos{2mz}.
    \end{eqnarray}
\end{subequations}
Self-advection by the $\epsilon$-order solution gives rise to no correction to vertical velocity $w_2=0$, a correction to horizontal velocity $(u_2,\,v_2)$ and vorticity $q_2$ fields that is $z$-invariant, and temperature field that is horizontally invariant \citep[Fig.~\ref{fig:mechanism_diagram}\emph{b,c}, also shown in][]{Boubnov_and_Golitsyn_1995}. Meanwhile, the NOB effects lead to a $w_2$ correction with wavenumber $(k_h,\ 2m)$ (red arrow).
As illustrated in Fig.~\ref{fig:mechanism_diagram}(\emph{d}), this correction enhances convection in the lower column while diminishing it in the upper column ($\epsilon$-order solution shown by gray contours). This symmetry breaking in the vertical direction arises from the fact that a fluid parcel with a fixed temperature anomaly gains more buoyancy in the lower part of the domain due to the amplified thermal expansivity. This amplification acts equally on the upward and downward plumes.

At $\epsilon^3$-order, the interaction between the $\epsilon^2$-order solution on one hand, and the $\epsilon$-order solution (black arrows) or the NOB term (red arrows) on the other, calls for corrections at five different wavenumbers. Among them, the forcing with wavenumber $(k_h,\,m)$ leads to a constraint on the amplitude of the $\epsilon$-order solution following the solvability condition (Fig.~\ref{fig:mechanism_diagram}\emph{e}, also see Sec.~\ref{subsec:amplitude}), whilst the rest leads to the following correction of $\epsilon^3$-order.
\begin{eqnarray}\label{eq:w3}
    w_3&=& W_{31}\sin{(kx+ly)}\sin{3mz}+W_{32}\sin{3(kx+ly)}\sin{mz} \nonumber \\
    && +\cos{2(kx+ly)}(W_{33}\sin{mz}+W_{34}\sin{3mz}).
\end{eqnarray}
The two correction terms with a horizontal wavenumber of $2k_h$ lead to a symmetry breaking between the upward and downward plumes (Fig.~\ref{fig:mechanism_diagram}\emph{h}). As to be shown later (Sec.~\ref{subsec:asymmetry}), whether upward or downward plumes are more dominant depends on $T_a$ and $P_r$. Similar horizontal asymmetry align with results from previous studies \citep{Horn_et_al_2013,Buijs_2015,Liu_et_al_2018}.
Detailed derivation and formulas of coefficients are given in Appendix~\ref{sec:appA}.

\begin{figure*}
    \centering
    \includegraphics[width=\textwidth]{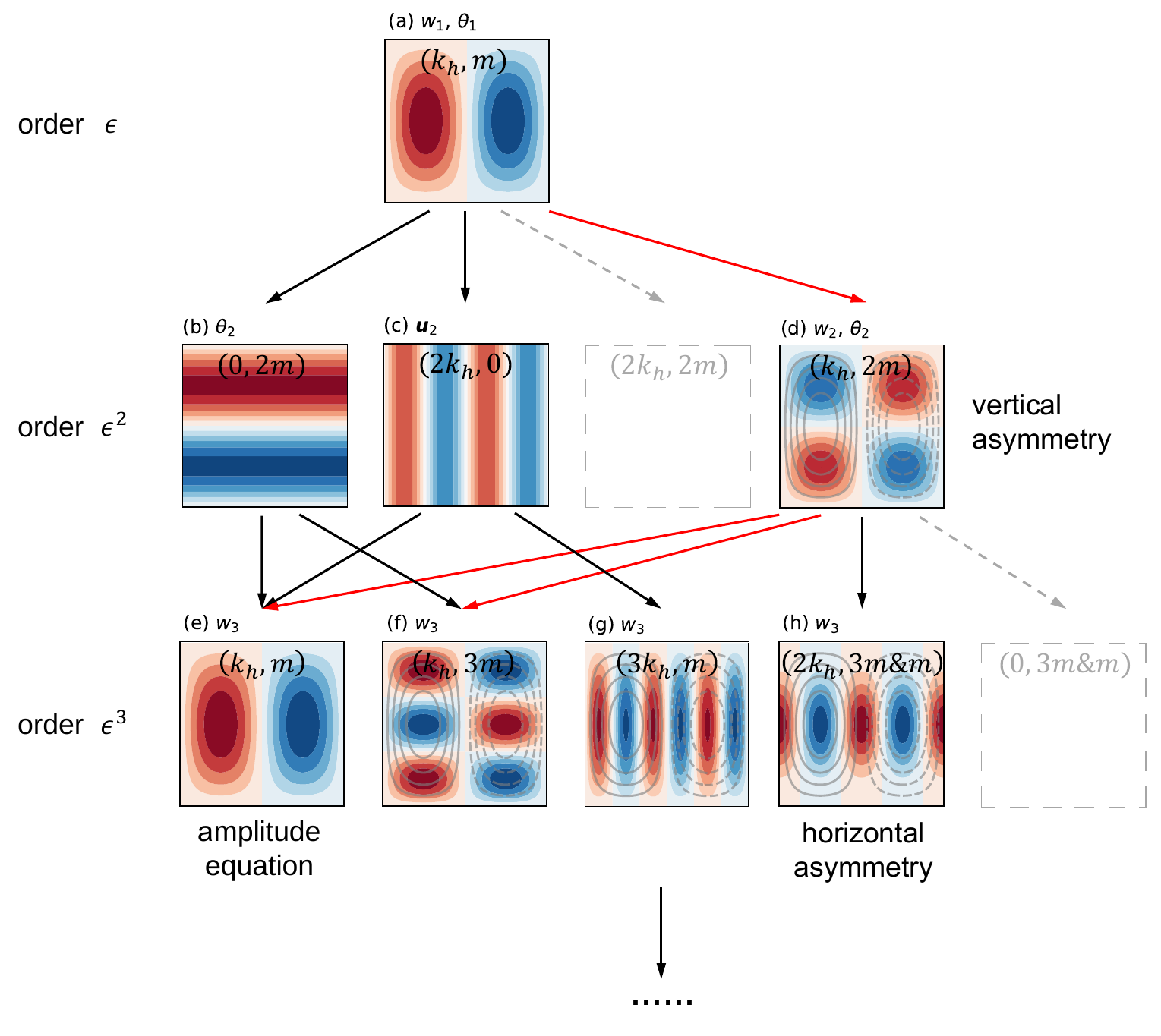}
    \caption{A schematic diagram of asymptotic expansion. Color shadings represent (\emph{a}) $w_1$ and $\theta_1$ at $\epsilon$-order, (\emph{b--c}) $\theta_2$, $\bf u_2$, and $w_2$ at $\epsilon^2$-order, and (\emph{e--h}) $w_3$ at $\epsilon^3$-order, respectively. In each panel, the wavenumber is listed. Gray contour lines in panel (\emph{d--h}) represent the solution in panel (\emph{a}). Arrows represent the generation of higher-order components, black ones are directly related to nonlinear advection, and red ones are directly related to NOB buoyancy. Empty boxes and gray dashed arrows mark the vanished components owing to boundary conditions. Panels (\emph{f--h}) are obtained with $T_a=10^3$ and $P_r=7$.}
    \label{fig:mechanism_diagram}
\end{figure*}


\subsection{Amplitude equation}\label{subsec:amplitude}
The resonance term at $\epsilon^3$-order leads to the following amplitude equation, by imposing the solvability condition.
\begin{equation}\label{eq:amplitude_equation}
    g_0\frac{\partial W_1}{\partial T}=g_1W_1-g_3W_1^3,
\end{equation}
where 
\begin{subequations}
    \begin{eqnarray}
        g_0&=&\frac{\epsilon^{-2}}{k_h^2/m^2+1}\left(1+\frac{1}{Pr}\frac{K_{11}^6-T_am^2}{K_{11}^6+T_am^2}\right), \\
    g_1&=&m^2\left(\frac{1}{\epsilon^2}\frac{\delta R_a}{R_{ac}}+\frac{K_{11}^2}{2K_{12}^2}\chi_2\right), \\
    g_3&=&\frac{1}{8(k_h^2/m^2+1)}\left(1-\frac{m^2}{k_h^2P_r^2}\frac{T_am^2}{K_{11}^6+T_am^2}\right),    
    \end{eqnarray}
\end{subequations}
are coefficients as functions of $P_r$ and $T_a$ (wavenumbers are determined by $T_a$ in Eq.~\ref{eq:critical_1st_wavenumber}). Here $K_{12}^2=k_h^2+4m^2$, and $\chi_{2}$ is a positive dimensionless coefficient. The derivation and the forms of coefficients are shown in Appendix~\ref{sec:appA}. 

\begin{figure*}
    \centering
    \includegraphics[width=\textwidth]{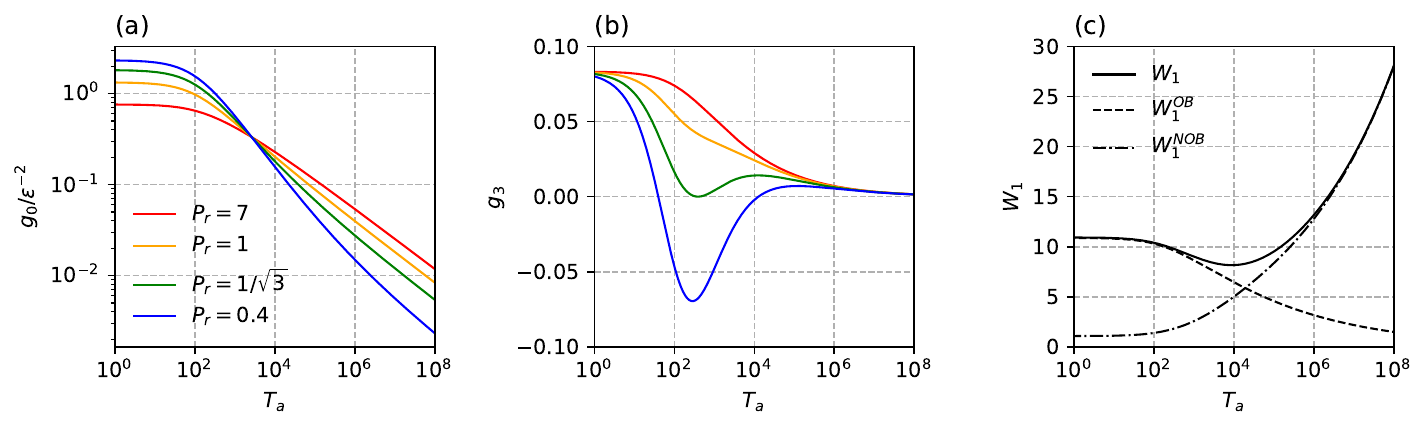}
    \caption{(\emph{a} \& \emph{b}) Functional relations between $T_a$ and $g_0$ scaled by $\epsilon^{-2}$ and $g_3$ for different values of $P_r$, respectively. (\emph{c}) Functional relations between $T_a$ and $W_1$ (solid curve), $W_1^{\rm OB}$ (dashed curve), and $W_1^{\rm NOB}$ (dash-dot curve) at $P_r=7$.}
    \label{fig:coef_g0_g3}
\end{figure*}

The dependencies of $g_0$ and $g_3$ on $T_a$ and $P_r$ are illustrated in Fig.~\ref{fig:coef_g0_g3}(\emph{a,b}). Our results are generally consistent with \cite{Scheel_2007}, except that the magnitudes of $g_0$ and $g_3$ are off by a constant factor. This discrepancy may stem from two key distinctions: firstly, they applied the no-slip condition at top and bottom plates and employed hyperbolic functions in the $z$-direction, whereas the present paper uses the stress-free condition and sinusoidal functions; secondly, they considered slow changes along the axis aligned with convective rolls, whereas no changes in this direction here. For fluid with large $P_r$, such as water, the pitchfork bifurcation is always supercritical and stable ($g_3>0$). However, in fluids with very small $P_r$, there exist a range of $T_a$ where the bifurcation is subcritical but unstable ($g_3<0$), corresponding to an intermittent convection, as proposed by \cite{Bajaj_et_al_2002} and \cite{Scheel_2007}. \cite{Frohlich_et_al_1992} and \cite{Liu_et_al_2018} found that pronounced $\hat{\nu},\,\hat{\kappa}$-related NOB effects can also cause this subcritical pitchfork bifurcation.  


When the convection reaches equilibrium state (i.e., $\frac{\partial W_1}{\partial T}=0$), the final amplitude satisfies
\begin{equation}\label{eq:amplitude_equation_steady}
    W_1^2=\underbrace{\frac{m^2}{g_3\epsilon^2}\frac{\delta R_a}{R_{ac}}}_{(W_1^{\rm OB})^2}+\underbrace{\frac{m^2K_{11}^2}{2g_3K_{12}^2}\chi_2}_{(W_1^{\rm NOB})^2}
\end{equation}
The first term corresponds to the $w$ amplitude under OB approximation, and the second term is an additional component related to NOB effects. When there are no NOB effects and rotation, Eq.~(\ref{eq:amplitude_equation_steady}) reduces to $8K_{11}^2\delta R_a/(\epsilon^2R_{ac})$, consistent with standard results under OB approximation \citep{Yano_2023_chap4}. 

In Fig.~\ref{fig:coef_g0_g3}(\emph{c}), we show the magnitude of convective velocity as a function of $T_a$ (Eq.~\ref{eq:amplitude_equation_steady}) with $P_r=7$. As we vary $T_a$, we remain $\delta R_a=\epsilon^2 R_{ac}^0$ as a constant, where $R_{ac}^0$ is the critical Rayleigh number for non-rotating RBC. By so doing, we consider a system whose Rayleigh number exceed critical values by a fixed amount regardless of the rotation rate. As can be seen, $\delta R_a/R_{ac}$ in Eq.~(\ref{eq:amplitude_equation_steady}) decreases with $T_a$, as $R_{ac}$ increases. The amplitude first slightly decreases then increases with $T_a$ due to the competition between a decreased $W_1^{\rm OB}$ and an increased $W_1^{\rm NOB}$. The decreased $W_1^{\rm OB}$ is consistent with the fact that strong rotation suppresses vertical motion, known as the Taylor-column effect \citep{Taylor_1922}. $W_1^{\rm NOB}$ is enhanced because the temperature anomaly in the interior increases with $T_a$ (e.g., Eq.~\ref{eq:solution_steady_1st_theta}), leading to a stronger NOB buoyancy forcing. These effects of rotation on $W_1^{\rm OB}$ and temperature anomaly are consistent with the numerical findings presented by \cite{Kunnen_et_al_2006}.

In fact, the $R_a$, where the bifurcation happens, does not equal to the critical $R_{ac}$ we obtained from linear analysis (Eq.~\ref{eq:critical_Rayleigh_number}) due to NOB effects. The following correction is therefore needed
\begin{equation}\label{eq:correction_Rac}
    R_{ac}^*=R_{ac}\left(1-\epsilon^2\frac{\chi_2K_{11}^2}{2K_{12}^2}\right),
\end{equation}
A similar $\epsilon^2$-correction has also been proposed by \cite{Paolucci_and_Chenoweth_1987}, \cite{Frohlich_et_al_1992}, \cite{Liu_et_al_2018}, and \cite{Pan_et_al_2023}, but to be able to suppress convection (i.e., increase $R_{ac}$) when the $\hat{\nu},\hat{\kappa}$-related NOB effects are considered. 

\subsection{Two types of symmetry breaking}\label{subsec:asymmetry}  
\begin{figure*}
    \centering
    \includegraphics[width=\textwidth]{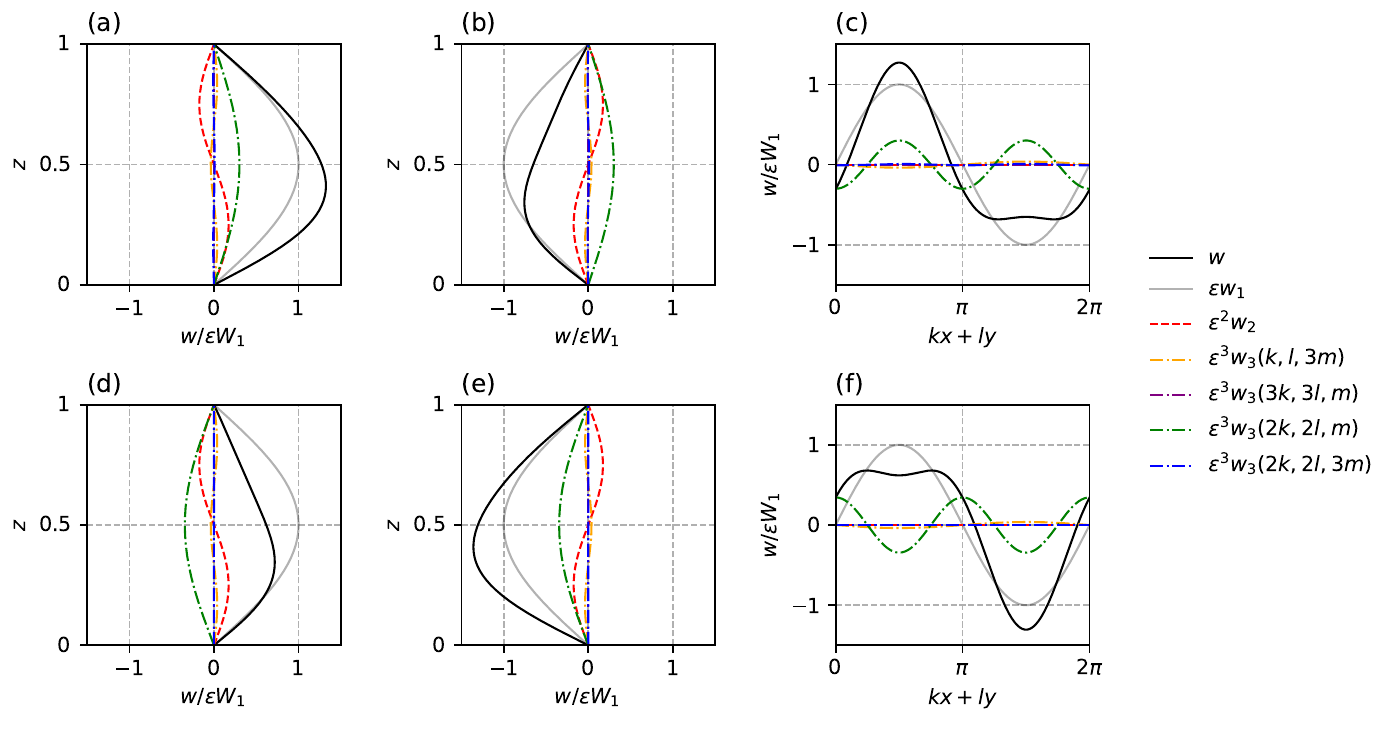}
    \caption{Structures of vertical velocity in (\emph{a} \& \emph{d}) the upwelling center, i.e., $kx+ly=\pi/2$, (\emph{b} \& \emph{e}) the downwelling center, i.e., $kx+ly=3\pi/2$, and (\emph{c} \& \emph{f}) the mid-plane, i.e., $z=0.5$, scaled by $\epsilon W_1$. Different colored curves represent components for each power of $\epsilon$. Black solid curves represent their sum. Panels (\emph{a--c}) are for $P_r=2$, and (\emph{d--e}) are for $P_r=7$. In all panels, $T_a=0$ and $\epsilon=3$.}
    \label{fig:total_w}
\end{figure*}

Fig.~\ref{fig:total_w} shows solution for $w$ with $P_r=2$ (panels \emph{a--c}) and $7$ (panels \emph{d--f}) in non-rotating convection ($T_a=0$). We set $\epsilon$ to 3 to emphasize NOB effects, which only show up in higher-order terms. Although this choice gets beyond where weakly nonlinear analysis should hold, the results still illuminate the correct general trends.

The key feature that shows up in Fig.~\ref{fig:total_w} is the asymmetry about the horizontal mid-plane, $w(x,\,y,\,z)\neq w(x,\,y,\,1-z)$. The vertical profiles of $w$ are bottom-heavy in both the upwelling and downwelling branches (panels \emph{a,b,d,e}). The most substantial contribution comes from the NOB-induced $w_2$ at $\epsilon^2$-order, as summarized in Fig. \ref{fig:mechanism_diagram}(\emph{d}). Similar asymmetry is also evident in other physical properties and is in line with previous literature on the NOB effects induced by the inhomogeneity of $\hat{\nu}$ and $\hat{\kappa}$ \citep[e.g.,][]{Frohlich_et_al_1992,Liu_et_al_2018}. However, a notable difference between the $\hat{\nu},\hat{\kappa}$-related and $\hat{\alpha}$-related NOB effects here is that, in the weakly nonlinear analysis, there is no asymmetry in the vertical profile of mean temperature or a shift of the center (bulk) temperature of the fluid. This difference stems from the fact that the $\hat{\alpha}$ variations in our work is prescribed to be a function of depth (pressure), whereas the variations of $\hat{\nu}$ and $\hat{\kappa}$ in previous work are set by temperature, which in turn is determined by dynamics. Our $\hat{\alpha}$ variations act equally on both the upward warm branch and the downward cold branch, thereby entirely offsetting any warming in the upward branch with the cooling in the downward branch.

\begin{figure}
    \centering
    \includegraphics[width=0.47\textwidth]{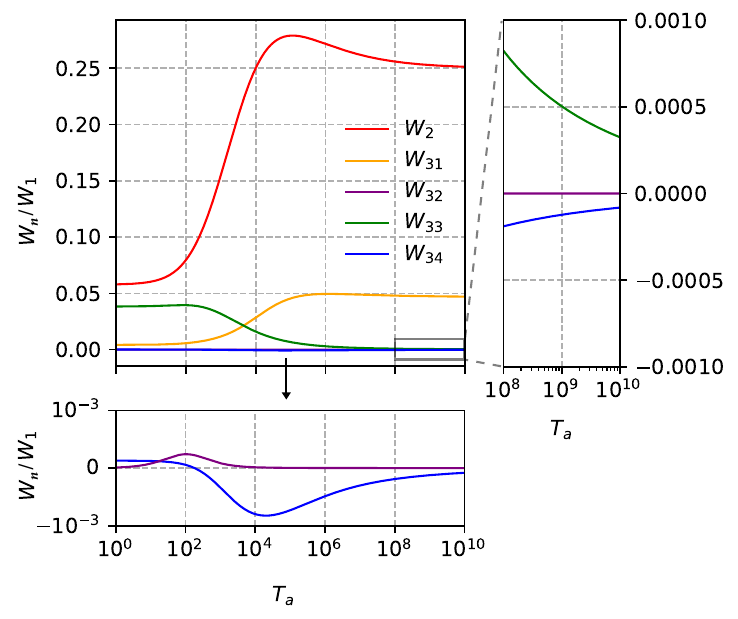}
    \caption{Functional relationship between $T_a$ and the coefficients of vertical velocity at each power of $\epsilon$ to $W_1$ (see Appendix~\ref{sec:appA}). $W_{32}$ (purple curve) and $W_{34}$ (blue curve) are very small so that individually displayed in the lower panel. Curves in the grey box are amplified in the right panel. $P_r$ is set to $7$.}
    \label{fig:coef_cn}
\end{figure}

Fig.~\ref{fig:total_w}(\emph{a,b,d,e}) also illustrates a symmetry breaking between upward and downward convection, i.e., $w(x,\,y,\,z)\neq-w(-x,\,-y,\,z)$. At $P_r=2$, upward convection is stronger than downward convection (panels \emph{a,b}), whilst downward convection is stronger at $P_r=7$ (panels \emph{d,e}). Fig.~\ref{fig:total_w}(\emph{c,f}) depicts the vertical velocity structure in the horizontal mid-plane $z=0.5$, highlighting the horizontal asymmetry. At $P_r=2$, the downward convection is weaker, more homogeneous and occupies a larger portion of the area, whilst the upward convection is stronger and more concentrated (panel \emph{c}). This phenomenon is primarily attributed to the fact that an asymmetric component of $w_3$ at $\epsilon^3$-order, which is with wavenumber of $(2k,\,2l,\,m)$, remains positive in both the upwelling and downwelling branches, arising from the interference between nonlinear advection and NOB-induced bottom-heavy profiles  (Fig.~\ref{fig:mechanism_diagram}\emph{h}). At $P_r=7$, the component of $w_3$ with wavenumber $(2k,\,2l,\,m)$ becomes negative in both braches of convection, leading to opposite asymmetry (Fig.~\ref{fig:total_w}\emph{f}). The sign of this asymmetry component, i.e., $W_{33}$ in Eq.~(\ref{eq:w3}), is determined by $T_a$ and $P_r$ (Eqs.~\ref{eq:G1G2_rotation} and \ref{eq:G1G2_nonrotation}). In the absence of rotation ($T_a=0$), upward convection predominates when $P_r$ is small, while downward convection becomes stronger when $P_r$ is large. The threshold for no horizontal asymmetry occurs at $P_r=3$. When rotation is strong ($T_a$ is very large), downward convection is consistently dominant.

Rotation can also affect the magnitudes of the two types of asymmetry, manifested in that $w$ at each order vary with $T_a$. In Fig.~\ref{fig:coef_cn}, we present the coefficients of $w$ at each order to $W_1$ as the function of $T_a$ (i.e., $W_n/W_1$, obtained from Appendix~\ref{sec:appA}). These ratios represent the relative strengths of $w$ at each power of $\epsilon$ to $w_1$ at $\epsilon=1$. The increase in $W_2/W_1$ with $T_a$ indicates that the contribution from $w_2$ becomes more prominent as rotation gets stronger, i.e., the vertical asymmetry is more obvious. It mainly arises from the NOB buoyancy (see forcing terms Eq.~\ref{eq:term_NOB_direct}), and thus is expected to strengthen as the temperature anomaly increases with rotation. In contrast, the magnitudes of $W_{33}/W_1$ and $W_{34}/W_1$ decrease with rotation and approach zero as $T_a$ increases, suggesting that stronger rotation weakens the horizontal asymmetry. These two components are primarily generated by the advection of temperature, vorticity, and divergence (see forcing terms Eqs.~\ref{eq:term_temp_advection}--\ref{eq:term_divergence_advection}). However, the stronger rotation makes the flows follow the thermal-wind relation more closely, meaning that the flows align almost parallel to the isotherms and become less effective in advecting heat, vorticity, and divergence. As a result, the corresponding vertical motion weakens.

\section{Numerical simulations}\label{sec:numerical_results}
\subsection{Two types of symmetry breaking}
In this section, we use numerical simulations to validate our asymptotic solutions. As demonstrated in previous sections, the $\hat{\alpha}$-induced NOB effects mainly have two effects. First, the $\epsilon^2$-order correction has a vertical wavenumber of $2m$, making the convection bottom-heavy. The degree of this vertical asymmetry is quantified by a ratio $r_1$
\begin{equation}\label{eq:ratio_w2/w1}
    r_1=\frac{\epsilon^2W_2}{\epsilon W_1}=\epsilon\chi_2,
\end{equation}
where we utilize the expression of $W_2$ provided in Appendix~\ref{sec:appA}. Second, at $\epsilon^3$-order, the correction terms with the horizontal wavenumber of $2k_h$ induce an asymmetry between upward and downward plumes. This horizontal asymmetry primarily stems from $w_3$ with the wavenumber of $(2k,\,2l,\,m)$, and can be quantified by
\begin{equation}\label{eq:ratio_w3/w1}
    r_2=-\frac{\epsilon^3W_{33}}{\epsilon W_1}=c_r\epsilon\left(\frac{\delta R_a}{R_{ac}}+\epsilon^2\frac{K_{11}^2}{2K_{12}^2}\chi_2\right)^{1/2},
\end{equation}
where $c_r$ is a dimensionless coefficient varying with $T_a$ and $P_r$, and its specific value at arbitrary $T_a$ and $P_r$ can be determined using the expression of $W_{33}$ provided in Appendix~\ref{sec:appA}. Positive and negative values of $r_2$ suggest stronger upward and downward plumes, respectively. The magnitudes of $r_1$ and $r_2$ both increase with $\epsilon$, i.e., the strength of NOB effects, as expected. Meanwhile, they are modulated by $T_a$. Also, the magnitude of $r_2$, the measure of the asymmetry between upward and downward plumes, increases with supercriticality ($\delta R_a/R_{ac}$), which has been assumed to be $\epsilon^2$-order in our weakly nonlinear analysis.

We use the Dedalus solver \citep{Dedalus_ref} to integrate the dimensionless forms of governing equations for RBC (Eqs.~\ref{eq:general_du/dt}--\ref{eq:general_mass_conserve}) with the free-slip boundary conditions (Eq.~\ref{eq:boundary_condition}). The model is configured to have only two dimensions, $x$ and $z$, where all fields including $v$, the velocity in $y$-direction, remains unchanged along $y$-direction. By limiting the dynamics to 2D, we are able to conduct a decent number of simulations with relatively low computational cost. Our domain is periodic in the horizontal dimension(s) and has a dimensionless width of $2\sqrt{2}$. The vertical extension of our domain is set to $1$ (dimensionless). This width is chosen to accommodate the most unstable mode $k_h=m/\sqrt{2}$ and to mitigate the effects of finite domain on the most unstable $k_h$. The grid points are $128\times64$ ($x\times z$). 

\begin{figure*}
    \centering
    \includegraphics[width=\textwidth]{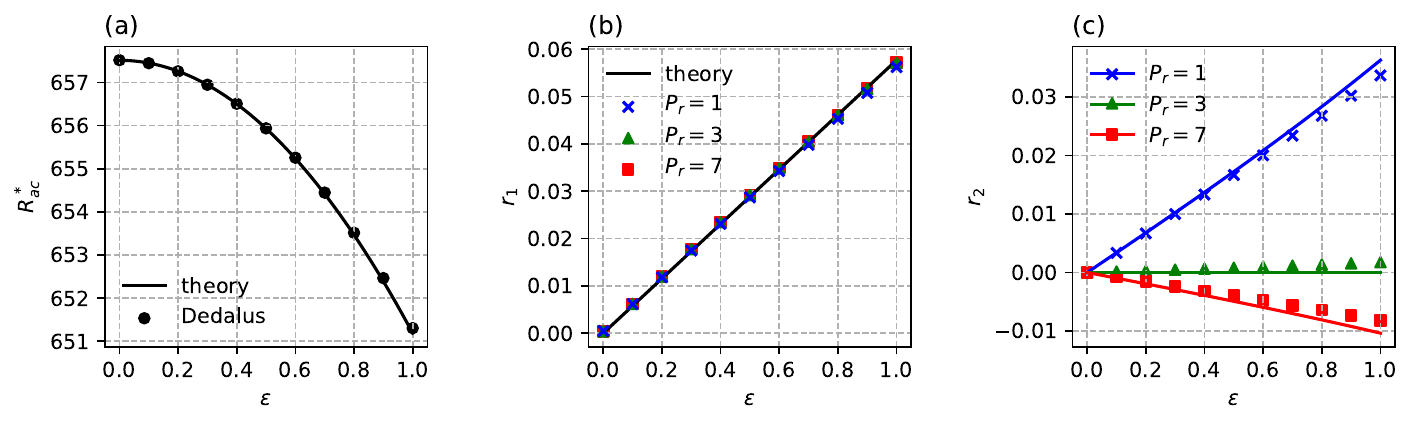}
    \caption{Dependencies of (\emph{a}) $R_{ac}^*$, (\emph{b}) $r_1$, and  (\emph{c}) $r_2$ on $\epsilon$. In panels (\emph{b}) and (\emph{c}), $R_a=700$, and $P_r=1$, $3$, $7$ (blue cross, green triangle, and red square, respectively). Theoretical predictions, Eqs. (\ref{eq:correction_Rac})--(\ref{eq:ratio_w3/w1}), are represented by the solid curves in each panel, respectively.}
    \label{fig:2d_dedalus_vs_theory}
\end{figure*}

We first explore how the bottom-heavy structure ($r_1$) and the asymmetry between the upward and downward plumes ($r_2$) vary with $\epsilon$ in the absence of rotation ($T_a=0$). The critical Rayleigh numbers are firstly examined. As illustrated in Fig.~\ref{fig:2d_dedalus_vs_theory}(\emph{a}), the critical Rayleigh number decreases with $\epsilon$ following a quadratic shape, indicating that NOB effects indeed promote convection onset at $\epsilon^2$-order, as predicted by Eq.~(\ref{eq:correction_Rac}). 

Knowing that $R_{ac}$ is around 650, we set a series of experiments with $R_a=700$ to ensure low supercriticality ($\delta R_a/R_{ac}<0.1$), and increase $\epsilon$ from 0 to 1 by an interval of 0.1. We also set $P_r=1$, $3$, and $7$ to verify different regimes of horizontal asymmetry mentioned in Sec.~\ref{subsec:asymmetry}.  The numerical solutions for these experiments (not shown) exhibit similar convection pattern to what is shown in Fig.~\ref{fig:mechanism_diagram}. As $\epsilon$ increased, the convection in the lower domain intensifies, and the vertical asymmetry becomes more pronounced, consistent with previous theoretical analyses. To quantify this asymmetry, we utilize Eq.~(\ref{eq:ratio_w2/w1}), with the $W_1$ and $W_2$ amplitudes measured by
\begin{eqnarray}
    \widetilde{W}_1&=&2\int_0^1w_{\rm c}\sin{\pi z}\mathrm{d}z, \\
    \widetilde{W}_2&=&2\int_0^1w_{\rm c}\sin{2\pi z}\mathrm{d}z,
\end{eqnarray}
where $w_{\rm c}$ is the vertical velocity magnitude profile measured at centers of plumes, both upward and downward. Fig.~\ref{fig:2d_dedalus_vs_theory}(\emph{b}) shows that $r_1$ from numerical experiments increases from zero in the OB case to approximately $6\%$, regardless of $P_r$. This matches almost perfectly with the prediction by Eq.~(\ref{eq:ratio_w2/w1}) (shown by black curve). 

To quantify the asymmetry between upward and downward plumes ($r_2$), we measure $W_{33}$ from numerical simulations as,
\begin{equation}\label{eq:decompose_W33}
    \widetilde{W}_{33}=\frac{w_{\rm up}^{\rm mid}+w_{\rm down}^{\rm mid}}{2},
\end{equation}
where the superscript `mid' represents the $w$ profile in the mid-plane $z=0.5$, and subscripts `up' and `down' denote upward and downward centers, respectively. Fig.~\ref{fig:2d_dedalus_vs_theory}(\emph{c}) illustrates the ratio $r_2=\widetilde{W}_{33}/\widetilde{W}_1$ at various $\epsilon$ and $P_r$. When $P_r=1$, $r_2$ is positive and increases with $\epsilon$, indicating that stronger NOB effects lead to larger horizontal asymmetry, with upward plumes stronger than downward ones. In contrast, $r_2$ becomes negative at $P_r=7$, signifying that downward plumes are more dominant. A marginal scenario with nearly zero horizontal asymmetry is observed at $P_r=3$, where $r_2$ in the experiments approaches zero. These results align with the predictions in Sec.~\ref{subsec:asymmetry} and Eq.~(\ref{eq:ratio_w3/w1}). It is worth noting that the power law of $r_2$ with respect to $\epsilon$ is nearly linear rather than quadratic. This is because the bracketed term in Eq.~(\ref{eq:ratio_w3/w1}) is dominated by the constant $\delta R_a/R_{ac}$, which is about $6\%$ and is much larger than the maximum NOB correction (about $1\%$ at $\epsilon=1$). 

Previous studies have observed similar symmetry breaking in the velocity field, although quantitative characterization has been lacking. \cite{Horn_et_al_2013} simulated NOB convection in glycerol and revealed asymmetries in the velocity field. Similarly,  \cite{Buijs_2015} conducted direct numerical simulations of water convection, identifying an asymmetric structure in the vertical velocity field, possibly due to variations in $\hat{\alpha}$ in the experiments.

\setlength{\tabcolsep}{10pt}
\begin{table*}\normalsize
    \centering
    \renewcommand{\arraystretch}{1.5}
    \begin{tabular}{l c c c c c c c c c}
    \hline\hline
        $\frac{k_h}{m}$ & $\frac{1}{\sqrt{2}}$ & $\frac{1}{\sqrt{2}}$ &$\frac{2}{\sqrt{2}}$ & $\frac{3}{\sqrt{2}}$ & $\frac{4}{\sqrt{2}}$ & $\frac{5}{\sqrt{2}}$ & $\frac{6}{\sqrt{2}}$ & $\frac{7}{\sqrt{2}}$ & $\frac{8}{\sqrt{2}}$\\ \hline
        $T_a$ & 1 & 10 &2630 & 23573 & 118352 & 426068 & 1230764 & 3040333 & 6682946 \\ \hline
        $R_{ac}$ & 658 & 677 & 2630 & 8840 & 23671 & 53259 & 105494 & 190021 & 318236 \\ \hline
        $R_a$ & 700 & 700 &2672 & 8882 & 23713 & 53301 & 105536 & 190063 & 318278 \\ \hline\hline
    \end{tabular}
    \caption{Most unstable wavenumbers ($\frac{k_h}{m}$), Taylor numbers ($T_a$), critical Rayleigh numbers in OB case ($R_{ac}$), and the setting Rayleigh numbers ($R_a$). $T_a$ is calculated from Eq.~(\ref{eq:critical_1st_wavenumber}) and $R_{ac}$ is calculated from Eq.~(\ref{eq:critical_Rayleigh_number}).}
    \label{tab:Ra_vs_Ta}
\end{table*}

\begin{figure*}
    \centering
    \includegraphics[width=\textwidth]{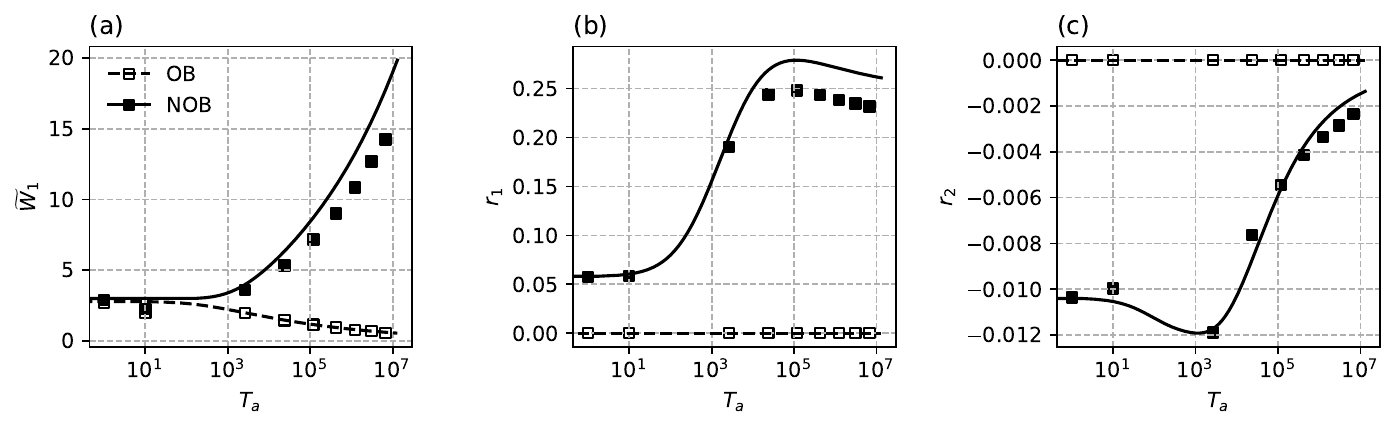}
    \caption{Dependencies of (\emph{a}) $\widetilde{W}_1$, (\emph{b}) $r_1$, and (\emph{c}) $r_2$ on $T_a$. Empty and solid squares represent OB and NOB ($\epsilon=1$) experiments. Theoretical predictions (Eqs.~\ref{eq:amplitude_equation_steady}, \ref{eq:ratio_w2/w1}, and \ref{eq:ratio_w3/w1}) for OB and NOB scenarios are represented by the dashed and solid curves in each panel, respectively.}
    \label{fig:3d_dedalus_vs_theory}
\end{figure*}

We then explore how rotation affects the two types of asymmetry ($r_1$ and $r_2$) by taking $P_r=7$ as an example. $T_a$ and $R_a$ used in our experiments are summarized in Table~\ref{tab:Ra_vs_Ta}. These $R_a$ values are chosen to remain $\delta R_a$ constant ($=43$) to be consistent with the assumption in Sec.~\ref{subsec:asymmetry}. Also, these $T_a$ values are calculated from our theory to ensure that there are integer numbers of convection cells of the gravest mode in the domain. Since the supercriticality is low here ($\delta R_a/R_{ac}<0.1$), we expect the most unstable mode to match our theoretical prediction. The results are shown in Fig.~\ref{fig:3d_dedalus_vs_theory}. The strength of the convection (panel \emph{a}), quantified by $\widetilde{W}_1$, decreases as the system's rotation rate ($T_a$) increases in the OB cases \citep{Kunnen_et_al_2006}, and in contrast, convection strengthens with increasing rotation rate in the NOB cases, in line with Eq.~(\ref{eq:amplitude_equation_steady}) and Fig.~\ref{fig:coef_g0_g3}(\emph{c}). The strengthening of $\widetilde{W}_1$ stems from the enhanced NOB buoyancy as rotation rate increases. In Fig.~\ref{fig:3d_dedalus_vs_theory}(\emph{b}), the vertical asymmetry $r_1$, which also arises from NOB buoyancy, monotonically increases with $T_a$, agreeing with the theoretical prediction by Eq.~(\ref{eq:ratio_w2/w1}) as illustrated by the $W_{2}/W_1$-curve in Fig.~\ref{fig:coef_cn}. This asymmetry reaches up to $25\%$ at large $T_a$. The asymmetry between upward and downward plumes, as measured by the magnitude of $r_2$, generally decreases with $T_a$ (Fig.~\ref{fig:3d_dedalus_vs_theory}\emph{c}). This decrease is because stronger rotation forces the dynamics to follow thermal-wind relation more closely, which weakens heat/vorticity advection and hence horizontal asymmetry, as discussed in Sec.~\ref{subsec:asymmetry}. Although the magnitude of $r_2$ in NOB experiments is only less than $1\%$, the signals can still be measured, and they match well with the prediction given by Eq.~(\ref{eq:ratio_w3/w1}), which is illustrated by the $W_{33}/W_1$-curve in Fig.~\ref{fig:coef_cn}.

\subsection{Heat transport efficiency}
\begin{figure*}
    \centering
    \includegraphics[width=\textwidth]{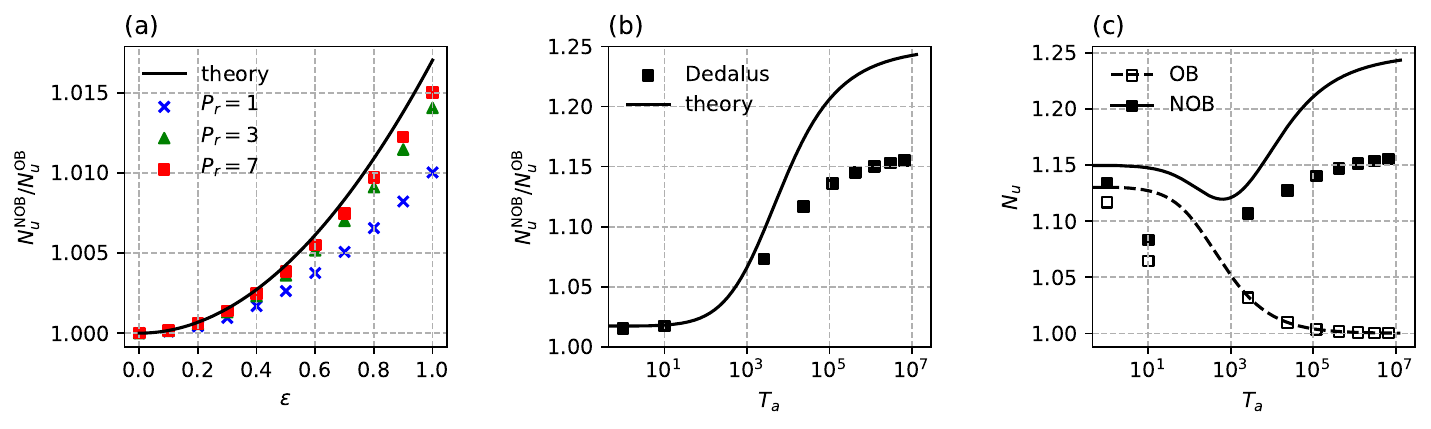}
    \caption{(\emph{a}) Dependence of the ratio $N_u^{\rm NOB}/N_u^{\rm OB}$ on $\epsilon$ and (\emph{b}) $T_a$, and (\emph{c}) dependence of $N_u$ on $T_a$ in OB (empty squares) and OB convection (solid squares). In panel (\emph{a}), blue crosses, green triangles, and red squares denote Dedalus results at $P_r=1$, $3$, and $7$, respectively. In panel (\emph{a}), $T_a=0$. In panel (\emph{b}) and (\emph{c}), $\epsilon=1$ for NOB experiments, and $P_r=7$. Theoretical predilections (Eq.~\ref{eq:Nu}) are shown by black curves. In panel (\emph{c}), the pair of points at $T_a=10$ deviates a lot from the theory because they do not correspond to an integer number of convective cells.}
    \label{fig:Nu}
\end{figure*}

In this section, we explore the influences of NOB effects on heat transport efficiency, which is measured by the ratio of Nusselt number in NOB convection relative to that in OB equivalents, $N_u^{\rm NOB}/N_u^{\rm OB}$. In our asymptotic analysis, this ratio can be expressed analytically,
\begin{eqnarray}\label{eq:Nu}
    \frac{N_u^{\rm NOB}}{N_u^{\rm OB}}&=&\frac{\sum\limits_{i=0}^{3} \left(\sum\limits_{j=0}^{i}\overline{w_j\theta_{i-j}}-\frac{\mathrm{d}}{\mathrm{d}z}\overline{\theta_i}\right)^{\rm NOB}}{\sum\limits_{i=0}^{3} \left(\sum\limits_{j=0}^{i}\overline{w_j\theta_{i-j}}-\frac{\mathrm{d}}{\mathrm{d}z}\overline{\theta_i}\right)^{\rm OB}} \nonumber \\
    &=&\frac{1+\epsilon^2W_1^2/(4K_{11}^2)}{1+\epsilon^2(W_1^{\rm OB})^2/(4K_{11}^2)},
\end{eqnarray}
where $W_1$ and $W_1^{\rm OB}$ are shown in Eq.~(\ref{eq:amplitude_equation_steady}). Note that $\epsilon^2(W_1^{\rm OB})^2$ in the denominator is independent with $\epsilon$, the ratio $N_u^{\rm NOB}/N_u^{\rm OB}$ would therefore increase with $\epsilon$, suggesting that NOB effects favor upward heat transport, in line with \cite{Ahlers_et_al_2006} and \cite{Sameen_2009}. This amplification is caused by the enhancement of vertical velocity (Eq.~\ref{eq:amplitude_equation_steady}) by our $\hat{\alpha}$ variation. Shown in Fig.~\ref{fig:Nu}(\emph{a}) is the simulated $N_u^{\rm NOB}/N_u^{\rm OB}$ as a function of $\epsilon$, clearly showing the predicted increasing trend. 
Note that in Eq.~(\ref{eq:Nu}), $N_u^{\rm NOB}/N_u^{\rm OB}$ is independent with $P_r$ in non-rotating convection. However, the numerical results indicate that the heat transfer efficiency increases with $P_r$. This deviation in magnitude may stem from that we only consider the first three-order solution in Eq.~(\ref{eq:Nu}), whilst higher-order terms have a dependence on $P_r$.

Our predictions from Eq.~(\ref{eq:Nu}) also suggest that the NOB effects on heat transfer will increase with rotation, measured by $T_a$, because rotation amplifies the magnitude of vertical velocity in the presences of NOB effects and then more heat been transported upward, as discussed in Sec.~\ref{subsec:amplitude}. This main trend is also captured by the numerical simulations, as shown in Fig.~\ref{fig:Nu}(\emph{b}). At large $T_a$, the simulated $w$ and $\theta$ are both slightly smaller than predicted values due to the higher-order corrections (Fig.~\ref{fig:3d_dedalus_vs_theory}\emph{a}), leading to a rather significant underestimate of $N_u^{\rm NOB}$ and hence the ratio $N_u^{\rm NOB}/N_u^{\rm OB}$ (Fig.~\ref{fig:Nu}\emph{b,c}). It is worth noting that if we plot $N_u^{\rm NOB}$ and $N_u^{\rm OB}$ against $T_a$ separately instead of showing their ratio, the two Nusselt numbers exhibit opposite trend as $T_a$ increases beyond a certain threshold (Fig.~\ref{fig:Nu}\emph{c}), indicating that, in that regime, heat transport will be suppressed by rotation in the OB case, but will be enhanced in the NOB case.

\section{Summary and discussion}\label{sec:summary}
In this study, we explore non-Oberbeck--Boussinesq (NOB) effects on rotating Rayleigh--B\'{e}nard Convection (RBC) due to the depth-dependent (pressure-dependent) thermal expansivity $\hat{\alpha}$. The convection is considered to be two-dimensional. NOB corrections can be manifested in various aspects of the fluid dynamics, we here focus on the convection patterns, especially the asymmetry between the upward and downward plumes and the vertical structure of each plume.

Using asymptotic analysis, we first find that NOB effects amplify the vertical motions in the lower part of the domain relative to the upper part at $\epsilon^2$-order, as the buoyancy acceleration rate is higher toward the bottom due to the high thermal expansivity there. At the next order ($\epsilon^3$-order), this vertical asymmetry gives rise to an asymmetry between the upward and downward plumes, making downward plumes more focused for most configurations unless both $P_r$ and $T_a$ are small. We then validate our analytical predictions using numerical methods in the weakly nonlinear regime. Although the vertical and horizontal asymmetries are challenging to discern visually from numerical simulations, they can be quantified using two ratios, $r_1$ (Eq.~\ref{eq:ratio_w2/w1}) measuring the vertical asymmetry and $r_2$ (Eq.~\ref{eq:ratio_w3/w1}) measuring the horizontal asymmetry. The strength of the two asymmetries monotonically increase with $\epsilon$. The magnitude of $r_1$ increases with $T_a$, because stronger rotation intensifies NOB-related buoyancy and enhances vertical asymmetry. Conversely, the magnitude of $r_2$ decreases with $T_a$, because stronger rotation leads to more closely aligned flows with the thermal-wind relation, impeding the transport of heat, vertical vorticity, and horizontal divergence, which are key to the generation of asymmetry between upward and downward plumes (horizontal asymmetry). 

In this work, we focus on convection under 2D configuration, precluding the possibility of pattern formation. It has been shown that, in OB rotating RBC system, roll, square and hexagon plume can spontaneously arise in such system in specific parameter regime \citep{Veronis_1959}, and that NOB effects would favor the formation of hexagonal patterns \citep[e.g,][]{Busse_1967,Madguga_et_al_2006,Ahlers_et_al_2010}. When NOB effects are present, previous simulations reveal significant asymmetry between upward and downward plumes. To explain this symmetry breaking, one needs to study how NOB factors modify the amplitude equation and change the fixed point solution and its stability, which is left to future work.
Furthermore, here we present only results in the weakly nonlinear regime. The transition toward highly nonlinear regime is another topic worth looking into.

As the rotating convective plumes become bottom-heavy (vertical asymmetry) and dominated by upward or downward plumes (horizontal asymmetry), the transport of passive tracers can be greatly influenced. For instance, \cite{Kang_et_al_2022} considered the tracer transport by convection in the ocean of an icy satellite, Enceladus. At low salinity, tracers released from the seafloor are found to concentrate near the bottom due to the strong NOB effects, compared to the high salinity cases. The tracer transport timescale plays a key role in the habitability and detectability of life activity on Enceladus, and thus deserves more detailed analysis.

\begin{acknowledgments}
WK acknowledges support from startup funding. We thanks the Dedalus Project to release the Dedalus codes.
\end{acknowledgments}

\section*{AUTHOR DECLARATIONS}
\subsection*{Conflict of Interest}
The authors report no conflict of interest.

\section*{DATA AVAILABILITY}
The data that support the findings of this study are available from the corresponding author upon reasonable request.

\appendix
\section{Solutions of convection}\label{sec:appA}
The asymptotic solution involves multiple distinct wavenumbers. For simplicity, we define the following symbols:
\begin{subequations}
    \begin{eqnarray}
        K_{11}^2&=&k_h^2+m^2,\quad\,\,\,\, K_{12}^2=k_h^2+4m^2,\quad\,\,\,\,\, K_{13}^2=k_h^2+9m^2, \nonumber \\
    K_{21}^2&=&4k_h^2+m^2,\quad K_{23}^2=4k_h^2+9m^2,\quad K_{31}^2=9k_h^2+m^2. \nonumber
    \end{eqnarray}
\end{subequations}

\subsection{$\epsilon$-order}
At $\epsilon$-order, retaining corresponding terms in Eqs. (\ref{eq:general_du/dt})--(\ref{eq:general_dvort/dt}) yields
\begin{eqnarray}
    D_{\nu}{\bf u}_1&=&-\sqrt{T_a}{\bf k}\times{\bf u}_1-\nabla\phi_1+R_{ac}\theta_1{\bf k}, \label{eq:1st_du/dt} \\
    D_{\kappa}\theta_1&=&w_1, \label{eq:1st_dtheta/dt} \\
    \sigma_{1}&=&-\frac{\partial w_1}{\partial z}, \label{eq:1st_mass_conserv} \\
    D_{\nu}q_{1}&=&-\sqrt{T_a}\sigma_{1}. \label{eq:1st_dvort/dt}
\end{eqnarray}
We consider a convection occurring at slightly supercritical Rayleigh number, which is usually assumed to be a quasi-steady convection with very slow changes. With this assumption, we would find steady solutions of the governing equations, where $D_{\nu}$ and $D_{\kappa}\rightarrow-\nabla^2$. We consider 2D configuration, where $w_1$ has the form
\begin{equation}\label{eq:solution_steady_1st_w}
    w_1=W_1\sin{(kx+ly)}\sin{mz},
\end{equation}
since it satisfies the stress-free boundary conditions~(\ref{eq:boundary_condition}). To be consistent with this $w$ ansatz, the other fields should take the following forms following Eqs. (\ref{eq:1st_du/dt})--(\ref{eq:1st_dvort/dt}),
\begin{eqnarray}
    u_1&=&U_1\cos{(kx+ly)}\cos{mz}\nonumber\\
    &=&\frac{m(kK_{11}^2+l\sqrt{T_a})}{ K_{11}^2k_h^2}W_1\cos{(kx+ly)}\cos{mz}, \label{eq:solution_steady_1st_u} \\
    v_1&=&V_1\cos{(kx+ly)}\cos{mz}\nonumber \\
    &=&\frac{m(lK_{11}^2-k\sqrt{T_a})}{ K_{11}^2k_h^2}W_1\cos{(kx+ly)}\cos{mz}, \label{eq:solution_steady_1st_v} \\
    \theta_1&=&\Theta_1\sin{(kx+ly)}\sin{mz}\nonumber\\
    &=&\frac{W_1}{ K_{11}^2}\sin{(kx+ly)}\sin{mz}, \label{eq:solution_steady_1st_theta} \\
    q_{1}&=&Q_1\sin{(kx+ly)}\cos{mz}\nonumber\\
    &=&\frac{\sqrt{T_a}mW_1}{ K_{11}^2}\sin{(kx+ly)}\cos{mz}, \label{eq:solution_steady_1st_vort} \\
    \sigma_{1}&=&\Sigma_1\sin{(kx+ly)}\cos{mz}\nonumber\\
    &=&-mW_1\sin{(kx+ly)}\cos{mz}, \label{eq:solution_steady_1st_conv}
\end{eqnarray}
which automatically satisfies the boundary conditions (\ref{eq:boundary_condition}).

\subsection{$\epsilon^2$-order}
At $\epsilon^2$-order, Eqs. (\ref{eq:general_du/dt})--(\ref{eq:general_dvort/dt}) include the following terms
\begin{eqnarray}
    D_{\nu}{\bf u}_2&=&-\frac{1}{P_r}{\bf u}_1\cdot\nabla{\bf u}_1-\sqrt{T_a}{\bf k}\times{\bf u}_2-\nabla\phi_2+R_{ac}[\theta_2+\theta_1\cos{(\pi z)}]{\bf k}, \label{eq:2nd_du/dt} \\
    D_{\kappa}\theta_2&=&-{\bf u}_1\cdot\nabla\theta_1+w_2, \label{eq:2nd_dtheta/dt} \\
    \sigma_{2}&=&-\frac{\partial w_2}{\partial z}, \label{eq:2nd_mass_conserv} \\
    D_{\nu}q_{2}&=&-\sqrt{T_a}\sigma_{2}+\frac{1}{P_r}[\nabla\times(-{\bf u}_1\cdot\nabla{\bf u}_1)]\cdot{\bf k}, \label{eq:2nd_dvort/dt}
\end{eqnarray}
We then cancel variables following the same manner as we derive Eq.~(\ref{eq:general_rRBC_w}) to obtain a single equation for $w_2$, which is forced by the nonlinear interaction of the $\epsilon$-order solution
\begin{eqnarray}\label{eq:2nd_rRBC_w}
    && \left[D_{\kappa}\left(D_{\nu}^2\nabla^2+T_a\frac{\partial^2}{\partial z^2}\right)-R_{ac} D_{\nu}\nabla_h^2\right]w_2 \nonumber \\ 
    &=&R_{ac}D_{\kappa}D_{\nu}[\cos{(\pi z)}\nabla_h^2\theta_1]+R_{ac}D_{\nu}\nabla_h^2(-{\bf u}_{1}\cdot\nabla\theta_1)+\frac{1}{P_r}D_{\kappa}D_{\nu}\nabla_h^2(-{\bf u}_1\cdot\nabla w_1) \nonumber \\
    &&-\frac{\sqrt{T_a}}{P_r}D_{\kappa}\frac{\partial}{\partial z}\left[\nabla\times(-{\bf u}_1\cdot\nabla{\bf u}_1)\right]\cdot{\bf k}-\frac{1}{P_r}D_{\kappa}D_{\nu}\frac{\partial}{\partial z}\left[\nabla_h\cdot(-{\bf u}_1\cdot\nabla{\bf u}_1)\right].
\end{eqnarray} 
Substituting the $\epsilon$-order solutions (Eqs. \ref{eq:solution_steady_1st_w}--\ref{eq:solution_steady_1st_conv}), we obtain the following nonlinear terms associated with NOB effects and advection, which show up on the right-hand side of Eq.~(\ref{eq:2nd_rRBC_w}):
\begin{eqnarray}
    R_{ac}D_{\kappa}D_{\nu}[\cos{(\pi z)}\nabla_h^2\theta_1]&=&-\frac{R_{ac}k_h^2}{2}\frac{K_{12}^4}{K_{11}^2}W_1\sin{(kx+ly)}\sin{2mz}, \\
    -{\bf u}_1\cdot\nabla\theta_1&=&-\frac{mW_1^2}{2 K_{11}^2}\sin{2mz}, \\
    -{\bf u}_1\cdot\nabla w_1&=&-\frac{mW_1^2}{2}\sin{2mz}, \\
    -{\bf u}_1\cdot\nabla(u_1,\,v_1)&=&\left(kK_{11}^2+l\sqrt{T_a},\,l K_{11}^2-k\sqrt{T_a}\right)\frac{m^2W_1^2}{2 K_{11}^2k_h^2}\sin{2(kx+ly)}.
\end{eqnarray}
The first term is related to NOB effects, and it is the only term that contributes to $w_2$. The other terms related to nonlinear advection vanishes after taking the spatial derivatives requested by Eq. (\ref{eq:2nd_rRBC_w}). Substituting these into Eq. (\ref{eq:2nd_rRBC_w}), we obtain the $w_2$ solution
\begin{equation}\label{eq:solution_steady_2nd_w}
    w_2= W_2\sin{(kx+ly)}\sin{2mz}=\chi_{2}W_1\sin{(kx+ly)}\sin{2mz},
\end{equation}
where
\begin{equation}
    \chi_{2}= \frac{K_{11}^6+T_am^2}{2[(K_{12}^6-K_{11}^6)+3T_am^2]}\cdot\frac{K_{12}^2}{K_{11}^2},
\end{equation}
is a positive dimensionless coefficient. Substituting $w_2$ into Eqs.~(\ref{eq:2nd_du/dt})--(\ref{eq:2nd_dvort/dt}), we obtain the expressions for other fields
\begin{eqnarray}
    u_2&=& U_{21}\cos{(kx+ly)}\cos{2mz}+U_{22}\sin{2(kx+ly)} \nonumber \\
    &=&\frac{2m(kK_{12}^2+l\sqrt{T_a})}{ K_{12}^2k_h^2}\chi_{2} W_1\cos{(kx+ly)}\cos{2mz}+\frac{\sqrt{T_a}lm^2W_1^2}{8P_rK_{11}^2k_h^4}\sin{2(kx+ly)}, \label{eq:solution_steady_2nd_u} \\
    v_2&=& V_{21}\cos{(kx+ly)}\cos{2mz}+V_{22}\sin{2(kx+ly)} \nonumber \\
    &=&\frac{2m(lK_{12}^2-k\sqrt{T_a})}{ K_{12}^2k_h^2}\chi_{2} W_1\cos{(kx+ly)}\cos{2mz}-\frac{\sqrt{T_a}km^2W_1^2}{8P_rK_{11}^2k_h^4}\sin{2(kx+ly)}, \label{eq:solution_steady_2nd_v} \\
    \theta_2&=& \Theta_{21}\sin{(kx+ly)}\sin{2mz}+\Theta_{22}\sin{2mz} \nonumber \\
    &=&\frac{W_1}{ K_{12}^2}\chi_{2}\sin{(kx+ly)}\sin{2mz}-\frac{W_1^2}{8mK_{11}^2}\sin{2mz}, \label{eq:solution_steady_2nd_theta} \\
    q_{2}&=& Q_{21}\sin{(kx+ly)}\cos{2mz}+Q_{22}\cos{2(kx+ly)} \nonumber \\
    &=&\frac{2\sqrt{T_a}mW_1}{ K_{12}^2}\chi_{2}\sin{(kx+ly)}\cos{2mz}-\frac{\sqrt{T_a}m^2W_1^2}{4P_rK_{11}^2k_h^2}\cos{2(kx+ly)}, \label{eq:solution_steady_2nd_vort} \\
    \sigma_{2}&=&\Sigma_2\sin{(kx+ly)}\cos{2mz}=-2m\chi_{2}W_1\sin{(kx+ly)}\cos{2mz}, \label{eq:solution_steady_2nd_conv} 
\end{eqnarray}
which automatically satisfies the boundary conditions (\ref{eq:boundary_condition}).

\subsection{$\epsilon^3$-order}
At $\epsilon^3$-order, retaining corresponding terms in Eqs. (\ref{eq:general_du/dt})--(\ref{eq:general_dvort/dt}) yields: 
\begin{eqnarray}
    D_{\nu}{\bf u}_3&=&-\frac{1}{P_r}({\bf u}_1\cdot\nabla{\bf u}_2+{\bf u}_2\cdot\nabla{\bf u}_1)-\sqrt{T_a}{\bf k}\times{\bf u}_3-\nabla\phi_3 \nonumber \\
    &&+R_{ac}[\theta_3+\theta_2\cos{(\pi z)}]{\bf k}-\frac{\partial{\bf u}_1}{P_r\epsilon^2\partial T}, \label{eq:3rd_du/dt} \\
    D_{\kappa}\theta_3&=&-{\bf u}_1\cdot\nabla\theta_2-{\bf u}_2\cdot\nabla\theta_1+w_3-\frac{1}{\epsilon^2}w_1\delta_{\Delta\theta}-\frac{\partial\theta_1}{\epsilon^2\partial T}, \label{eq:3rd_dtheta/dt} \\
    \sigma_{3}&=&-\frac{\partial w_3}{\partial z}, \label{eq:3rd_mass_conserv} \\
    D_{\nu}q_{3}&=&-\sqrt{T_a}\sigma_{3}+\frac{1}{P_r}[\nabla\times(-{\bf u}_1\cdot\nabla{\bf u}_2-{\bf u}_2\cdot\nabla{\bf u}_1)]\cdot{\bf k}-\frac{\partial q_{1}}{P_r\epsilon^2\partial T}. \label{eq:3rd_dvort/dt}
\end{eqnarray}
Here, the supercritical term and slowly changing terms occur, as the supercritical temperature lapse rate $\delta_{\Delta\theta}$ and $\partial/\partial T$ are $\epsilon^2$-order and act on $\epsilon$-order fields. Following the same 
procedure as in $\epsilon^2$-order, we obtain an equation for $w_3$ that is forced by the nonlinear interaction between $\epsilon$-order and $\epsilon^2$-order solutions
\begin{eqnarray}\label{eq:3rd_rRBC_w}
    && \left[D_{\kappa}\left(D_{\nu}^2\nabla^2+T_a\frac{\partial^2}{\partial z^2}\right)-R_{ac} D_{\nu}\nabla_h^2\right]w_3 \nonumber \\
    &=&R_{ac}D_{\kappa}D_{\nu}[\cos{(\pi z)}\nabla_h^2\theta_2]+\frac{\delta R_a}{\epsilon^2}D_{\nu}\nabla_h^2w_1 \nonumber \\
    &&R_{ac}D_{\nu}\nabla_h^2(-{\bf u}_1\cdot\nabla\theta_2-{\bf u}_2\cdot\nabla\theta_1)\nonumber \nonumber \\
    &&+\frac{1}{P_r}D_{\kappa}D_{\nu}\nabla_h^2(-{\bf u}_1\cdot\nabla w_2-{\bf u}_2\cdot\nabla w_1) \nonumber \\
    &&-\frac{\sqrt{T_a}}{P_r}D_{\kappa}\frac{\partial}{\partial z}\left[\nabla\times(-{\bf u}_1\cdot\nabla{\bf u}_2-{\bf u}_2\cdot\nabla{\bf u}_1)\right]\cdot{\bf k} \nonumber \\
    &&-\frac{1}{P_r}D_{\kappa}D_{\nu}\frac{\partial}{\partial z}\left[\nabla_h\cdot(-{\bf u}_1\cdot\nabla{\bf u}_2-{\bf u}_2\cdot\nabla{\bf u}_1)\right] \nonumber \\
    &&-\frac{R_{ac}}{\epsilon^2}D_{\nu}\nabla_h^2\frac{\partial\theta_1}{\partial T} +\frac
    {1}{P_r\epsilon^2}D_{\kappa}\left(\sqrt{T_a}\frac{\partial^2q_{1}}{\partial z\partial T}+D_{\nu}\frac{\partial^2\sigma_{1}}{\partial z\partial T}-D_{\nu}\nabla_h^2\frac{\partial w_1}{\partial T}\right).
\end{eqnarray}
Substituting $\epsilon$-order solutions (Eqs. \ref{eq:solution_steady_1st_w}--\ref{eq:solution_steady_1st_conv}) and $\epsilon^2$-order solutions (Eqs. \ref{eq:solution_steady_2nd_w}--\ref{eq:solution_steady_2nd_conv}), we can expand the right-hand side of Eq. (\ref{eq:3rd_rRBC_w}). These terms are categorized into a term directly arising from NOB effects:
\begin{eqnarray}\label{eq:term_NOB_direct}
    &&R_{ac}D_{\kappa}D_{\nu}[\cos{(\pi z)}\nabla_h^2\theta_2] \nonumber \\
    &=&-\frac{R_{ac}k_h^2}{2}\chi_{2}W_1\sin{(kx+ly)}\left(\frac{K_{11}^4}{K_{12}^2}\sin{mz}+\frac{K_{13}^4}{K_{12}^2}\sin{3mz}\right);
\end{eqnarray}
a supercritical term:
\begin{eqnarray}
    \frac{\delta R_a}{\epsilon^2} D_{\nu}\nabla_h^2w_1=-\frac
    {\delta R_a}{\epsilon^2}K_{11}^2k_h^2W_1\sin{(kx+ly)}\sin{mz};
\end{eqnarray}
a term arising from nonlinear advection of heat:
\begin{eqnarray}\label{eq:term_temp_advection}
    &&R_{ac}D_{\nu}\nabla_h^2(-{\bf u}_1\cdot\nabla\theta_2-{\bf u}_2\cdot\nabla\theta_1) \nonumber \\
    &=&-R_{ac}k_h^2mK_{21}^2\left(\frac{3}{K_{11}^2}-\frac{3}{K_{12}^2}\right)\chi_{2} W_1^2\cos{2(kx+ly)}\sin{mz} \nonumber \\
    &&+R_{ac}k_h^2mK_{23}^2\left(\frac{1}{K_{11}^2}-\frac{1}{K_{12}^2}\right)\chi_{2}W_1^2\cos{2(kx+ly)}\sin{3mz} \nonumber \\
    &&+\frac{R_{ac}k_h^2W_1^3}{8}\sin{(kx+ly)}\sin{mz} \nonumber \\
    &&-\frac{R_{ac}k_h^2W_1^3}{8}\frac{K_{13}^2}{K_{11}^2}\sin{(kx+ly)}\sin{3mz};
\end{eqnarray}
a term associated with the curl of nonlinear advection of horizontal momentum:
\begin{eqnarray}\label{eq:term_vorticity_advection}
    &&-\frac{\sqrt{T_a}}{P_r}D_{\kappa}\frac{\partial}{\partial z}\left[\nabla\times(-{\bf u}_1\cdot\nabla{\bf u}_2-{\bf u}_2\cdot\nabla{\bf u}_1)\right]\cdot{\bf k} \nonumber \\
    &=&-\frac{T_am^3}{2P_r}K_{21}^2\left(\frac{6}{K_{12}^2}+\frac{3}{K_{11}^2}\right)\chi_{2} W_1^2\cos{2(kx+ly)}\sin{mz} \nonumber \\
    &&+\frac{T_am^3}{2P_r}K_{23}^2\left(\frac{6}{K_{12}^2}-\frac{3}{K_{11}^2}\right)\chi_{2} W_1^2\cos{2(kx+ly)}\sin{3mz} \nonumber \\
    &&-\frac{T_am^4}{8P_r^2k_h^2}W_1^3\sin{(kx+ly)}\sin{mz} \nonumber \\
    &&-\frac{3T_am^4}{8P_r^2k_h^2}\frac{K_{31}^2}{K_{11}^2}W_1^3\sin{3(kx+ly)}\sin{mz};
\end{eqnarray}
a term associated with the divergence of nonlinear advection of horizontal momentum:  
\begin{eqnarray}\label{eq:term_divergence_advection}
    &&-\frac{1}{P_r}D_{\kappa}D_{\nu}\frac{\partial}{\partial z}\left[\nabla_h\cdot(-{\bf u}_1\cdot\nabla{\bf u}_2-{\bf u}_2\cdot\nabla{\bf u}_1)\right] \nonumber \\
    &=&+\frac{9m^3}{2P_r}K_{21}^4\chi_{2}W_1^2\cos{2(kx+ly)}\sin{mz} \nonumber \\
    &&-\frac{3m^3}{2P_r}K_{23}^4\chi_{2}W_1^2\cos{2(kx+ly)}\sin{3mz};
\end{eqnarray}
and the slowly changing terms:
\begin{eqnarray}
    &&-\frac{R_{ac}}{\epsilon^2}D_{\nu}\nabla_h^2\frac{\partial\theta_1}{\partial T}+\frac{1}{P_r\epsilon^2}D_{\kappa}\left(\sqrt{T_a}\frac{\partial^2q_{1}}{\partial z\partial T}+D_{\nu}\frac{\partial^2\sigma_{1}}{\partial z\partial T}-D_{\nu}\nabla_h^2\frac{\partial w_1}{\partial T}\right) \nonumber \\
    &=&\frac{1}{\epsilon^2}\left(\frac{P_r+1}{P_r}K_{11}^6+\frac{P_r-1}{P_r}T_am^2\right)\frac{\partial W_1}{\partial T}\sin{(kx+ly)}\sin{mz}.
\end{eqnarray}
Note that the nonlinear advection of vertical momentum is horizontally homogeneous, 
\begin{equation}
    -{\bf u}_1\cdot\nabla w_2-{\bf u}_2\cdot\nabla w_1=\frac{m}{2}\chi_{2}W_1^2(\sin{mz}-3\sin{3mz}),
\end{equation}
so the corresponding spatial derivative on the right-hand side of Eq. (\ref{eq:3rd_rRBC_w}) is zero. The above expressions suggest that the solutions of $w_3$ consists of components with wavenumbers of $(k,\,l,\,m)$, $(k,\,l,\,3m)$, $(2k,\,2l,\,m)$, $(2k,\,2l,\,3m)$, and $(3k,\,3l,\,m)$.

~\\
\begin{enumerate}
\item The wavenumber $(k,\,l,\,m)$ component falls into the null space of Eq. (\ref{eq:3rd_rRBC_w}). The solvability condition yields the following amplitude equation
\begin{eqnarray}
    &&\frac{1}{\epsilon^2}\left(\frac{P_r+1}{P_r}K_{11}^6+\frac{P_r-1}{P_r}T_am^2\right)\frac{\partial W_1}{\partial T} \nonumber \\
    &=&\frac{T_am^4}{8P_r^2k_h^2}W_1^3-\frac{R_{ac}k_h^2}{8}W_1^3+\frac{\delta R_a}{\epsilon^2}K_{11}^2k_h^2W_1+R_{ac}k_h^2\frac{\chi_{2}}{2}\frac{K_{11}^4}{K_{12}^2}W_1.
\end{eqnarray}
~\\

\item For wavenumber $(k,\,l,\,3m)$, the corresponding terms in Eq. (\ref{eq:3rd_rRBC_w}) will generate a vertical velocity with the same wavenumber: 
\begin{eqnarray}
    && \left[D_{\kappa}\left(D_{\nu}^2\nabla^2+T_a\frac{\partial^2}{\partial z^2}\right)-R_{ac} D_{\nu}\nabla_h^2\right]w_3 \nonumber \\
    &=&-\frac{R_{ac}k_h^2K_{13}^2}{2}\left(\frac{W_1^3}{4K_{11}^2}+\frac{K_{13}^2}{K_{12}^2}\chi_{2}W_1\right)\sin{(kx+ly)}\sin{3mz},
\end{eqnarray}
and the corresponding solution is
\begin{eqnarray}
    w_3&=&W_{31}\sin{(kx+ly)}\sin{3mz} \nonumber \\
    &=&\left(\frac{W_1^2}{4K_{11}^2}+\frac{K_{13}^2}{K_{12}^2}\chi_{2}\right)\chi_{31}W_1\sin{(kx+ly)}\sin{3mz},
\end{eqnarray}
where
\begin{equation}
    \chi_{31}= \frac{K_{11}^6+T_am^2}{2[(K_{13}^6-K_{11}^6)+8T_am^2]},
\end{equation}
is a positive dimensionless coefficient. This vertical velocity is produced by Eqs.~(\ref{eq:term_NOB_direct}) and (\ref{eq:term_temp_advection}), which are associated with direct NOB effects and heat advection, respectively.
~\\

\item For wavenumber $(3k,\,3l,\,m)$, the corresponding terms in Eq. (\ref{eq:3rd_rRBC_w}) will generate a vertical velocity with the same wavenumber:
\begin{eqnarray}
    && \left[D_{\kappa}\left(D_{\nu}^2\nabla^2+T_a\frac{\partial^2}{\partial z^2}\right)-R_{ac} D_{\nu}\nabla_h^2\right]w_3 \nonumber \\
    &=&-\frac{3T_am^4}{8P_r^2k_h^2}\frac{K_{31}^2}{K_{11}^2}W_1^3\sin{3(kx+ly)}\sin{mz},
\end{eqnarray}
and the correspongding solution is 
\begin{eqnarray}
    w_3&=&W_{32}\sin{3(kx+ly)}\sin{mz} \nonumber \\
    &=&\frac{3T_am^4}{4P_r^2K_{11}^2k_h^2(K_{11}^6+T_am^2)}\chi_{32}W_1^3\sin{3(kx+ly)}\sin{mz},
\end{eqnarray}
where 
\begin{equation}
    \chi_{32}= \frac{K_{11}^6+T_am^2}{2\left[(K_{31}^6-9K_{11}^6)-8T_am^2\right]},
\end{equation}
is a positive dimensionless factor, proven by eliminating $T_a$ via Eq. (\ref{eq:critical_1st_wavenumber}). This vertical velocity is produced by Eq.~(\ref{eq:term_vorticity_advection}), which is associated with vorticity advection.
~\\

\item For wavenumber $(2k,\,2l,\, m)$ and $(2k,\,2l,\, 3m)$, the corresponding terms in Eq. (\ref{eq:3rd_rRBC_w}) will generate one vertical velocity with vertical wavenumber $m$ and the other with vertical wavenumber $3m$, respectively. Together, they can be written as
\begin{eqnarray}\label{eq:3rd_rRBC_w_2kh}
    && \left[D_{\kappa}\left(D_{\nu}^2\nabla^2+T_a\frac{\partial^2}{\partial z^2}\right)-R_{ac} D_{\nu}\nabla_h^2\right]w_3 \nonumber \\
    &=&-\frac{T_am^3}{2P_r}K_{21}^2\left(\frac{6}{K_{12}^2}+\frac{3}{K_{11}^2}\right)\chi_{2} W_1^2\cos{2(kx+ly)}\sin{mz} \nonumber \\
    &&+\frac{9m^3}{2P_r}K_{21}^4\chi_{2} W_1^2\cos{2(kx+ly)}\sin{mz} \nonumber \\
    &&-R_{ac}k_h^2mK_{21}^2\left(\frac{3}{K_{11}^2}-\frac{3}{K_{12}^2}\right)\chi_{2} W_1^2\cos{2(kx+ly)}\sin{mz} \nonumber \\
    &&+\frac{T_am^3}{2P_r}K_{23}^2\left(\frac{6}{K_{12}^2}-\frac{3}{K_{11}^2}\right)\chi_{2} W_1^2\cos{2(kx+ly)}\sin{3mz} \nonumber \\
    &&-\frac{3m^3}{2P_r}K_{23}^4\chi_{2} W_1^2\cos{2(kx+ly)}\sin{3mz} \nonumber \\
    &&+R_{ac}k_h^2mK_{23}^2\left(\frac{1}{K_{11}^2}-\frac{1}{K_{12}^2}\right)\chi_{2} W_1^2\cos{2(kx+ly)}\sin{3mz}.
\end{eqnarray}
For simplifying solution, we use Eqs. (\ref{eq:critical_Rayleigh_number}) to eliminate $R_{ac}$ in the above equation. After doing so, we obtain the following solution:
\begin{eqnarray}\label{eq:solution_steady_3rd_w_2kh}
    w_3&=&\cos{2(kx+ly)}(W_{33}\sin{mz}+W_{34}\sin{3mz}) \nonumber \\
    &=&+\frac{9m^9}{P_rK_{11}^2K_{12}^2(K_{11}^6+T_am^2)}G_{1}\chi_{2}\chi_{33}W_1^2\cos{2(kx+ly)}\sin{mz} \nonumber \\
    &&-\frac{3m^9}{P_rK_{11}^2K_{12}^2(K_{11}^6+T_am^2)}G_{2}\chi_{2}\chi_{34}W_1^2\cos{2(kx+ly)}\sin{3mz}, 
\end{eqnarray}
where
\begin{subequations}
    \begin{eqnarray}
        \chi_{33}&=&\frac{ K_{11}^6+T_am^2}{2[(K_{21}^6-4K_{11}^6)-3T_am^2]}, \\
    \chi_{34}&=&\frac{ K_{11}^6+T_am^2}{2[(K_{23}^6-4K_{11}^6)+5T_am^2]}, \\
    G_{1}&=&\frac{2k_h^8}{m^8}+\left(6P_r+3\right)\frac{k_h^6}{m^6}+\left(12P_r-15\right)\frac{k_h^4}{m^4}+\left(6P_r-22\right)\frac{k_h^2}{m^2}-6, \\
    G_{2}&=&\frac{2k_h^8}{m^8}+\left(6P_r-5\right)\frac{k_h^6}{m^6}+\left(12P_r-35\right)\frac{k_h^4}{m^4}+\left(6P_r-62\right)\frac{k_h^2}{m^2}-34,    
    \end{eqnarray}
\end{subequations}
are four dimensionless coefficients. 

\end{enumerate}

The signs of the above four coefficients largely determine the spatial pattern of Eq.~(\ref{eq:solution_steady_3rd_w_2kh}), and we give a brief discussion here. We prove that both $\chi_{33}$ and $\chi_{34}$ are positive by eliminating $T_a$ using Eq.~(\ref{eq:critical_1st_wavenumber}), and $\chi_{33}$ is larger than $\chi_{34}$. However, the signs of $G_{1}$ and $G_{2}$ depend on $T_a$ and $P_r$. When $T_a$ is very large, we can estimate from Eq. (\ref{eq:critical_1st_wavenumber}) that $k_h$ is much larger than $m$. Consequently, this yields
\begin{equation}\label{eq:G1G2_rotation}
    G_1,\,\,G_2\rightarrow\frac{2k_h^8}{m^8},
\end{equation}
and they are both positive. This is so-called rotating-dominant scenario, where $w_3$ is mainly produced by Eq. (\ref{eq:term_vorticity_advection}), a term related to vorticity advection. In non-rotating RBC, $k_h=m/\sqrt{2}$, causing that 
\begin{eqnarray}\label{eq:G1G2_nonrotation}
    G_1=\frac{27}{4}(P_r-3),\quad G_2=\frac{27}{4}(P_r-11).
\end{eqnarray}
For the mixtures of gases that $P_r$ is very small, both $G_{1}$ and $G_{2}$ are negative, resulting in $w_3$ favoring upward motions. While for water that $P_r=7$, $G_{1}$ becomes positive, resulting in an opposite $w_3$ that favors downward motions. No matter which scenario, it is always true that $G_{1}$ is larger than $G_{2}$.

\nocite{*}
\bibliography{aipsamp_v1}

\end{document}